\documentclass[10pt,twocolumn,letterpaper]{article}

\usepackage{cvpr}              

\usepackage{graphicx}
\usepackage{amsmath}
\usepackage{amssymb}
\usepackage{booktabs}
\usepackage{bbm}
\usepackage{makecell}
\usepackage{multirow}
\usepackage{color}
\usepackage{url}

\usepackage[pagebackref,breaklinks,colorlinks]{hyperref}

\usepackage[capitalize]{cleveref}
\crefname{section}{Sec.}{Secs.}
\Crefname{section}{Section}{Sections}
\Crefname{table}{Table}{Tables}
\crefname{table}{Tab.}{Tabs.}

\def\cvprPaperID{8210} 
\def\confName{CVPR}
\def\confYear{2022}

\begin{document}

\title{Transformer-empowered Multi-scale Contextual Matching and Aggregation for \\ Multi-contrast MRI Super-resolution}

\author{Guangyuan Li$^1$, Jun Lv$^{1*}$, Yapeng Tian$^2$, Qi Dou$^3$, Chengyan Wang$^4$, Chenliang Xu$^2$, Jing Qin$^5$\\
$^1$School of Computer and Control Engineering, Yantai University, Yantai, China\\
$^2$University of Rochester\\
$^3$Department of Computer Science and Engineering, The Chinese University of Hong Kong, Hong Kong\\
$^4$Human Phenome Institute, Fudan University, Shanghai, China\\
$^5$Centre for Smart Health, School of Nursing, The Hong Kong Polytechnic University, Hong Kong\\
{\tt\small lgy1428275037@163.com, ljdream0710@pku.edu.cn,
$\{$yapengtian, chenliang.xu$\}$@rochester.edu,
} \\ 
{ \tt\small qidou@cuhk.edu.hk, wangcy@fudan.edu.cn, harry.qin@polyu.edu.hk} \\ 
}

\maketitle
\footnotetext{*Corresponding author. }

\begin{abstract}
Magnetic resonance imaging (MRI) can present multi-contrast images of the same anatomical structures, enabling multi-contrast super-resolution (SR) techniques. Compared with SR reconstruction using a single-contrast, multi-contrast SR reconstruction is promising to yield SR images with higher quality by leveraging diverse yet complementary information embedded in different imaging modalities. 
However, existing methods still have two shortcomings: (1) they neglect that the multi-contrast features at different scales contain different anatomical details and hence lack effective mechanisms to match and fuse these features for better reconstruction; and (2) they are still deficient in capturing long-range dependencies, which are essential for the regions with complicated anatomical structures.
%
We propose a novel network to comprehensively address these problems by developing a set of innovative Transformer-empowered multi-scale contextual matching and aggregation techniques; we call it \emph{McMRSR}. 
Firstly, we tame transformers to model long-range dependencies in both reference and target images. 
Then, a new multi-scale contextual matching method is proposed to capture corresponding contexts from reference features at different scales. 
Furthermore, we introduce a multi-scale aggregation mechanism to gradually and interactively aggregate multi-scale matched features for reconstructing the target SR MR image. 
Extensive experiments demonstrate that our network outperforms state-of-the-art approaches and has great potential to be applied in clinical practice.
Codes are available at \url{https://github.com/XAIMI-Lab/McMRSR}.

\end{abstract}

\begin{figure}[h]
	\centering
	\subfloat[Crop area]{\label{fig:a}\includegraphics[width=1in]{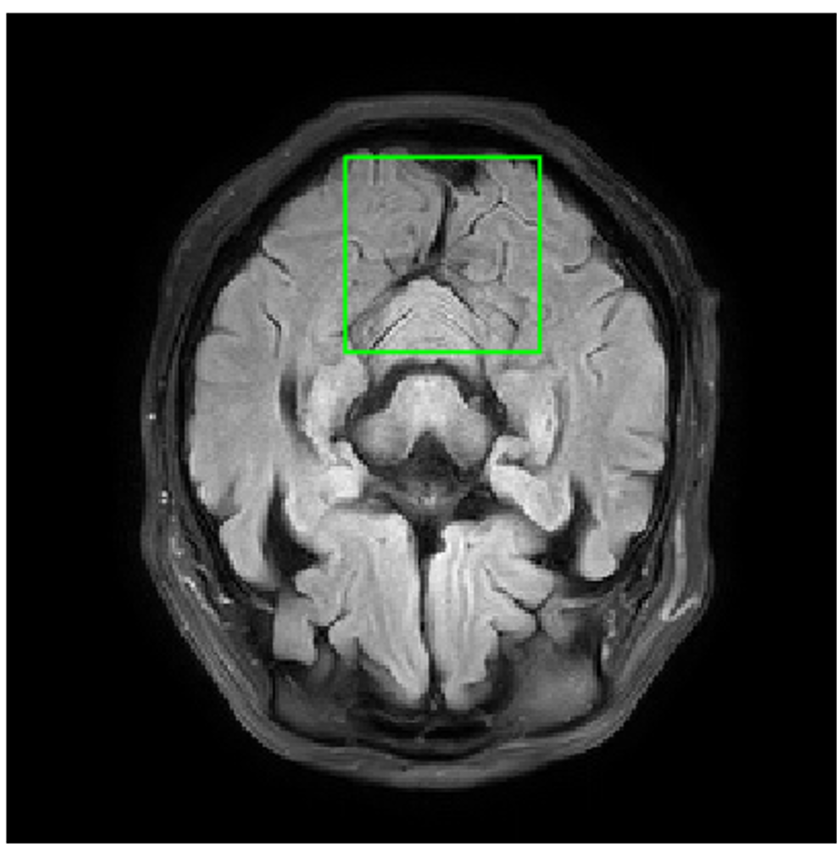}}\quad
	\subfloat[Bicubic]{\label{fig:b}\includegraphics[width=1in]{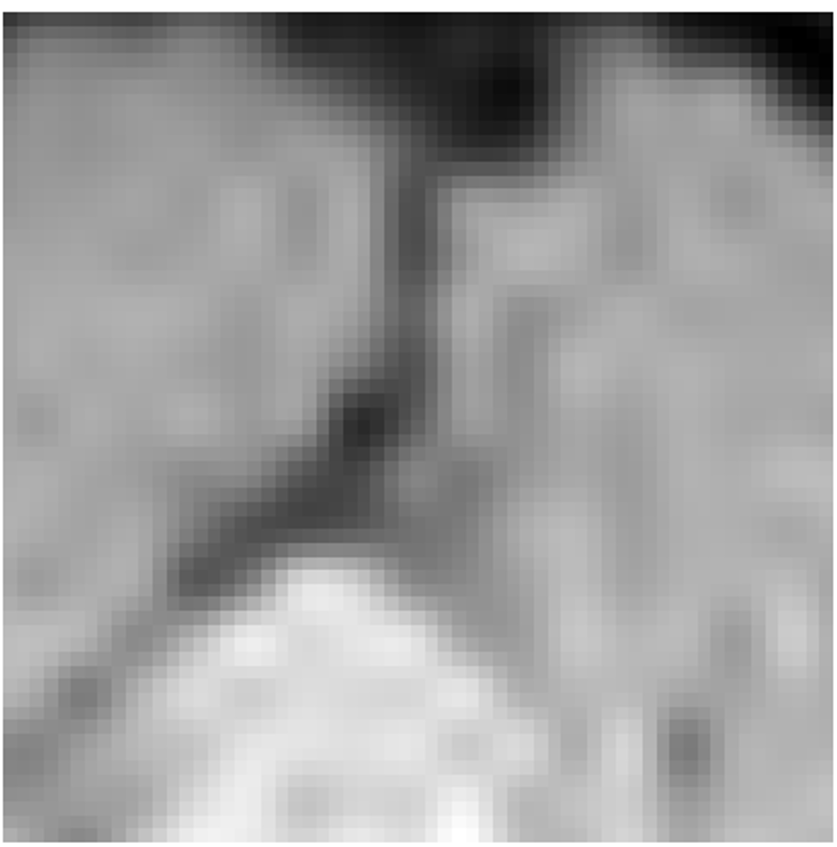}}\quad
	\subfloat[MCSR \cite{ lyu2020multi }]{\label{fig:c}\includegraphics[width=1in]{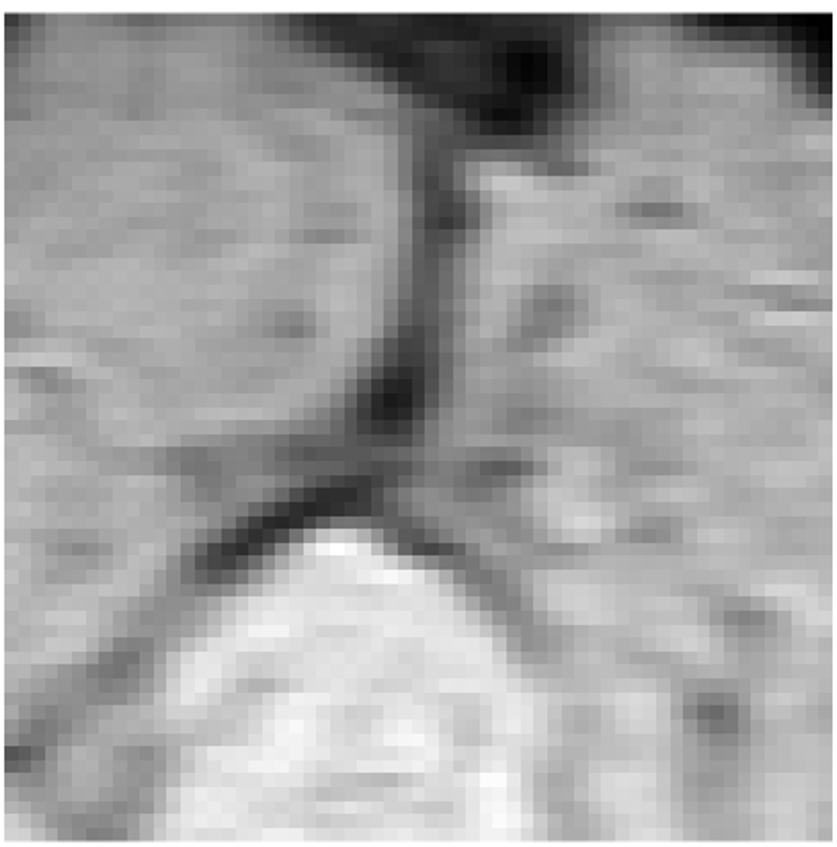}}\quad
	\subfloat[MINet \cite{ feng2021multi } ]{\label{fig:d}\includegraphics[width=1in]{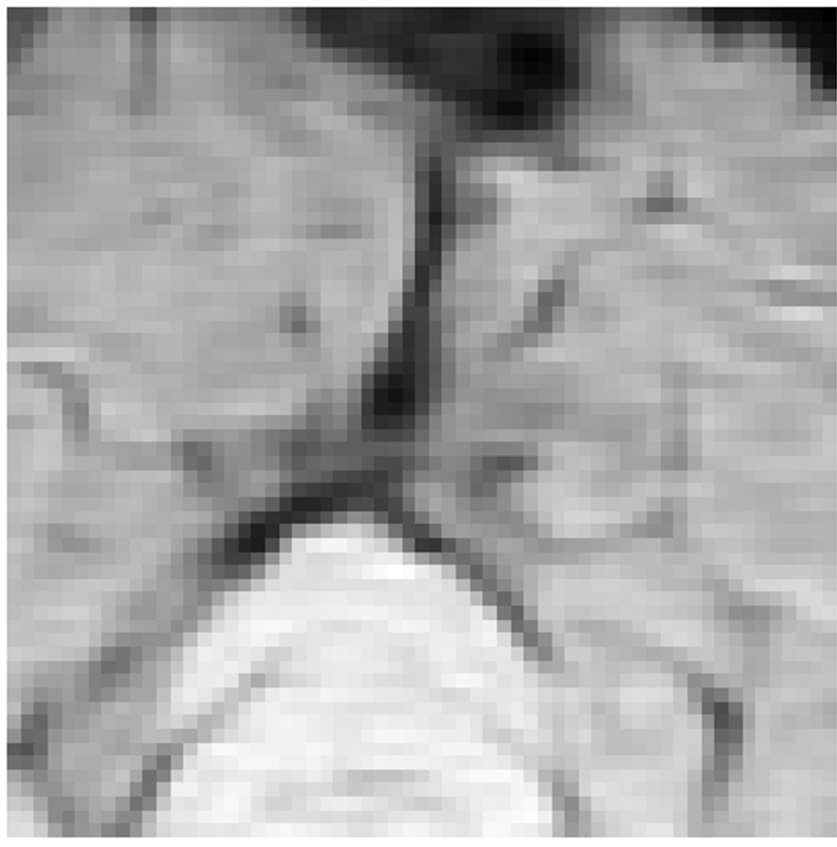}}\quad
	\subfloat[McMRSR (Ours)]{\label{fig:e}\includegraphics[width=1in]{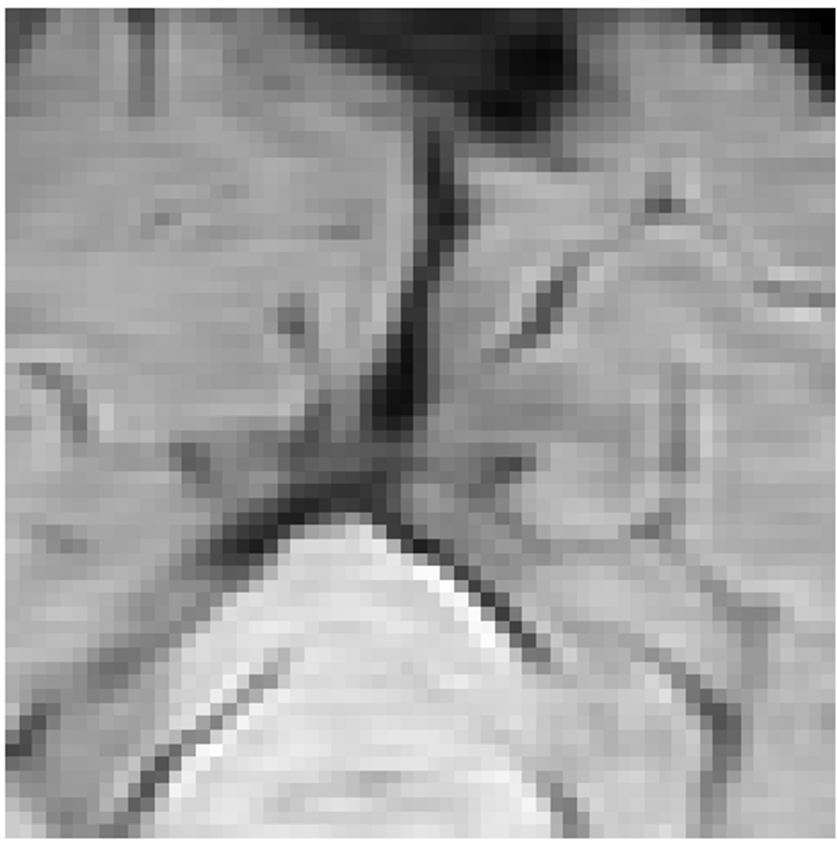}}\quad
	\subfloat[HR]{\label{fig:f}\includegraphics[width=1in]{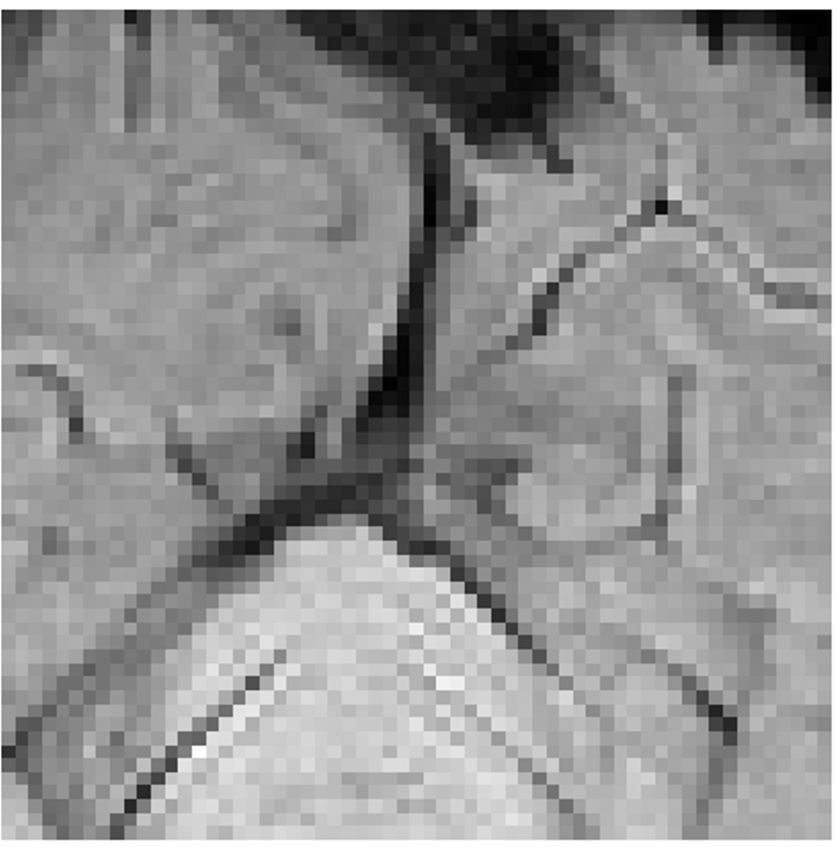}}\\	
	\vspace{-2mm}
	\caption{Compared with state-of-the-art multi-constrast MRI SR reconstruction methods: MCSR and MINet; the reconstructed MRI image by our McMRSR network contains sharper edges, more visual details, and fewer blurring artifacts.}
	\vspace{-5mm}
\end{figure}

\section{Introduction}
\label{sec:intro}
Magnetic resonance imaging (MRI) is an essential medical imaging technique in clinical application that provides clear information on tissue structure and function without causing ionizing radiation. 
However, due to the essential drawbacks of imaging systems \cite{plenge2012super, van2012super } and crepitations in some parts of the body, \eg, the abdomen, it is challenging to acquire high-resolution (HR) MR images in clinics~\cite{ feng2021brain }. 
In addition, prolonged acquisition procedure may cause discomforts to patients, introduce motion artifacts, and hence affect image quality \cite{ li2021high }. Super-resolution (SR) reconstruction is a promising way to improve the quality of MR images without upgrading hardware facilities \cite{ feng2021task }. 

MRI can present multi-contrast images with the same anatomical structures at different settings, \eg, T1-weighted images (T1) and T2-weighted images (T2), as well as proton density weighted images (PD) and fat-suppressed proton density weighted images (FS-PD), which can provide complementary information to each other~\cite{feng2021multi, chen2015accuracy}. 
In clinical applications, as the repetition time and echo time of T1 are shorter than those of T2 and the scanning process of PD is usually shorter than that of FS-PD, T1 can be used to guide LR T2 for SR reconstruction and PD can help to reconstruct FS-PD \cite{ xiang2018deep }. 
In this regard, it is promising to leverage an HR reference image with shorter acquisition time to reconstruct the modality with longer scanning time from an LR image. 

While some effort has been dedicated to multi-contrast MRI SR reconstruction \cite{feng2021multi, zheng2017multi, zeng2018simultaneous, zheng2018multi, stimpel2019multi, lyu2020multi }, we still face challenges in two key steps: (1) how to effectively extract the features in the reference and target images, and (2) how to transfer the features of the reference image to the features of the target image. 
In recent studies, Zeng \etal \cite{ zeng2018simultaneous } employed CNN to simultaneously perform single- and multi-contrast SR reconstruction. Lyu \etal \cite{ lyu2020multi } applied a GAN-based progressive network to multi-contrast SR reconstruction. Feng \etal \cite{ feng2021multi } used multi-stage integration network to perform multi-contrast MRI SR reconstruction. However, these methods are still incapable of sufficiently and comprehensively address the challenges in the two steps.

There are two main shortcomings.
First, most existing methods harness deep convolutional layers for feature extraction. 
However, the convolution kernel usually has a limited receptive field and hence cannot adequately capture long-range/non-local features, which are important for MRI SR reconstruction as, for some regions with complicated anatomical structures, faithful reconstruction depends on not only local relationships but also long-range dependencies.
Second, many of existing  methods~\cite{ lyu2020multi,feng2021multi} directly upsample the low-scale image into a high-scale image, and then perform the extraction and fusion of multi-contrast features.
However, these methods neglect that multi-contrast features at different scales contain different anatomical details and hence can provide broad yet diverse guidance for target MRI SR reconstruction.

In order to address these two shortcomings, in this paper, we propose a novel and effective network for multi-contrast MRI SR by taming transformers to extract long-range dependencies to facilitate more comprehensive contextual matching and exploiting multi-contrast multi-scale features to guide the reconstruction at different scales with anatomical information extracted from different modalities; we call the network as \emph{McMRSR} network.
%
%
Our contributions can be summarized as follows.
\begin{enumerate}
\item We propose a novel network equipped with transformer-empowered multi-scale contextual matching for multi-contrast MRI SR 
, where Swin Transformer groups are exploited to extract deep features at different scales and from different contrasts to capture more long-range dependencies.
\vspace{-2mm}
\item We propose multi-scale contextual matching and aggregation schemes to transfer visual contexts from reference images to target LR MR images at different scales, allowing the target LR images make full use of the guidance information to achieve SR images full of fine details.
\vspace{-2mm}


\item Our \emph{McMRSR} outperforms state-of-the-art approaches on three benchmark datasets: \emph{clinical pelvic}, \emph{clinical brain}, and \emph{fastMRI}, demonstrating its effectiveness and great potential to be used in clinical practice.
\end{enumerate}
\section{Related Work}
\label{sec:rela}
\subsection{Single-Contrast MRI SR}
 The commonly used interpolation methods \cite{ dong2015image } are bicubic and b-spline, but they introduce edge blurring and blocking artifacts in SR images, making it impossible for clinicians to make accurate diagnosis. Traditional SR algorithms exploit redundancy in the transform domain for MRI SR reconstruction, \emph{e.g.}, iterative deblurring algorithms \cite{ hardie2007fast , tourbier2015efficient } , low rank \cite{ shi2015lrtv } and dictionary learning \cite{ bhatia2014super }. However, when upsampling factor (UF) becomes large, the quality of the reconstructed SR images by these methods are not satisfactory. 
 
Following the research in deep learning-based natural image SR methods \cite{ dong2014learning, ledig2017photo, lim2017enhanced, mei2021image, zhang2021data, xing2021end} and computed tomography SR method \cite{yu2017computed}, some excellent MRI SR reconstruction methods emerged \cite{ oktay2016multi, mcdonagh2017context, chen2018brain, qiu2020super, lyu2020mri, steeden2020rapid, li2021high, du2020super, sun2020high, zhang2021mr}.  Qui \etal \cite{ qiu2020super } used a convolutional neural network (CNN) for knee MRI SR reconstruction. Lyu \etal \cite{ lyu2020mri } used ensemble learning for MRI SR reconstruction. Li \etal \cite{ li2021high } used attention mechanism and cyclic loss in GAN for pelvic image SR reconstruction. Zhang \etal \cite{ zhang2021mr } proposed squeezed and inspired inference attention network for MR image SR, and the experimental results showed the effectiveness of the method. However,
the above-mentioned algorithms all focus on reconstructing images by only using one contrast MR images. 

\begin{figure*}
  \centering
   \includegraphics[width=0.9\linewidth]{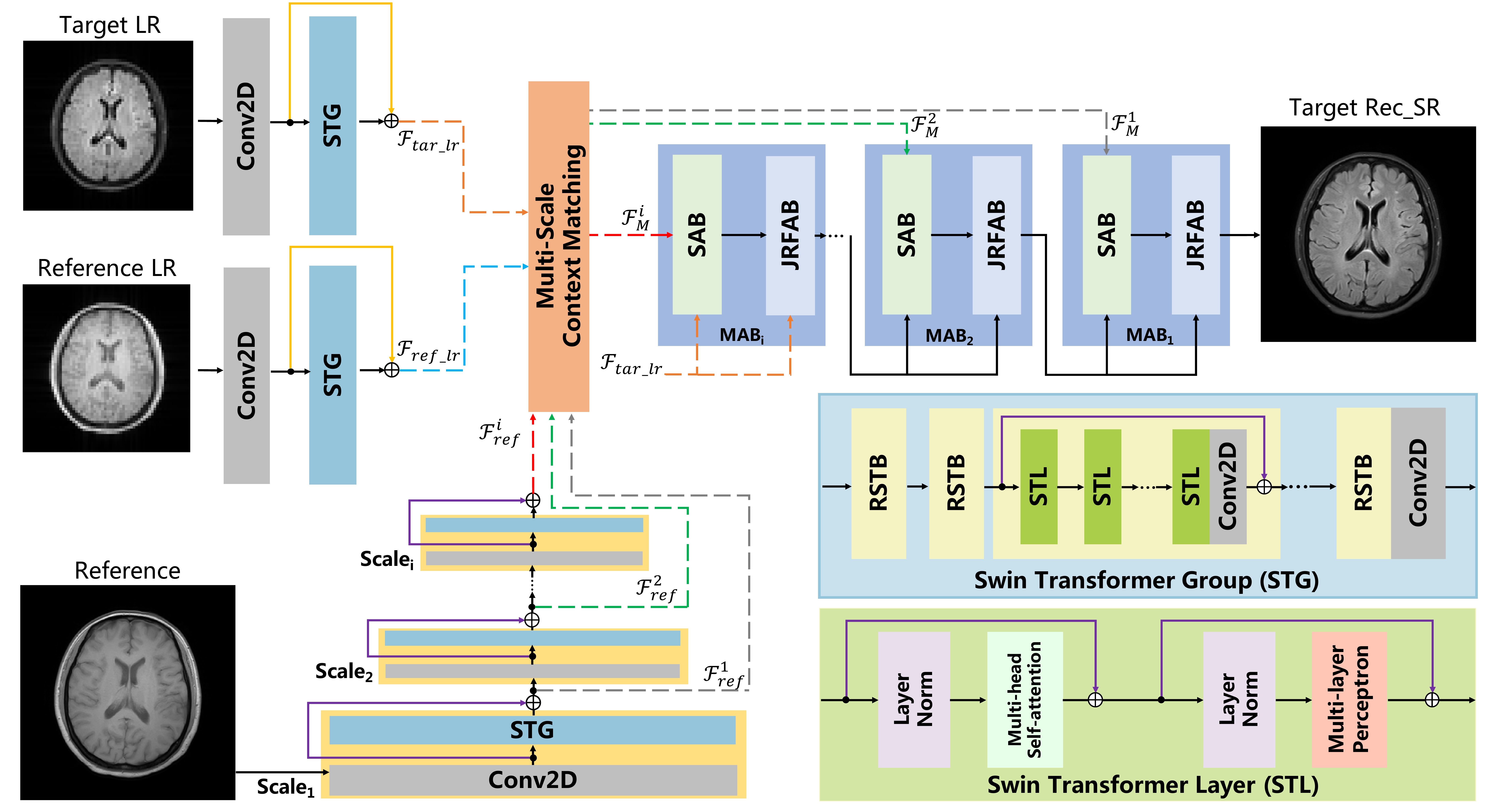}
   \caption{The overall architecture of the proposed McMRSR network. \textbf{STG}: Swin Transformer group; \textbf{RSTB}: residual Swin Transformer block; \textbf{STL}: Swin Transformer layer; \textbf{MAB}: multi-scale aggregation block; \textbf{SAB}: spatial adaptation block; \textbf{JRFAB}: joint residual feature aggregation block. 
   }
   \label{fig:model}
   \vspace{-5mm}
\end{figure*}
\subsection{Multi-Contrast MRI SR}
The key problem of the multi-contrast MRI SR is how to get the reference image to better guide the target image in SR reconstruction.
Lyu \etal \cite{ lyu2020multi } showed that the fusion of multi-contrast information in the high-level feature space yields better results than the combination in the low-level pixel space. Therefore, we consider multi-contrast feature matching and aggregation from the deep feature space to make full use of the information in the reference image. Feng \etal \cite{ feng2021multi } used a multi-stage feature fusion mechanism for multi-contrast SR, \emph{i.e.}, the reference features of the previous stage were fused with the target features to obtain the integrated features used to guide the learning of the target features in the next stage. Inspired by~\cite{ feng2021multi, lyu2020multi, lu2021masa}, we consider fusing features from reference images of different scales in the upsampling process. 
Concretely, we perform multi-scale context matching and aggregation in the deep feature space and use multi-scale matched reference features to guide the recovery of target HR features.

\subsection{MRI Transformer}
Unlike CNNs, the transformer \cite{ vaswani2017attention } uses a self-attentive mechanism to obtain global information between contexts and has achieved better results in dealing with visual problems \cite{ carion2020end, liu2021swin , touvron2021training }. In addition, there are several studies that have shown the effectiveness of transformer in MRI reconstruction. Feng \etal \cite{ feng2021task } used task transformer network to combine MRI reconstruction and SR reconstruction and proposed the use of multi-modal transformer for multi-contrast MRI reconstruction \cite{ feng2021accelerated }. However, the general transformer is processed in the form of image patches, which results in edge pixels not learning the information of neighboring pixels outside the patches \cite{ liang2021swinir }. Swin Transformer \cite{ liu2021swin } can be used to solve the above problem, which combines the advantages of CNN and general transformer. The method solves the problem of edge pixels in patch by shifting the window scheme to establish long-range dependencies \cite{ liang2021swinir }. Therefore, inspired by \cite{ liu2021swin, liang2021swinir }, we use Swin Transformer groups consisting of multiple residual Swin Transformer blocks in McMRSR for deep feature extraction and multi-contrast features fusion.

\section{Methodology}
\label{sec:method}

\subsection{Overll Architecture}
The overall architecture of the proposed McMRSR network is shown in \cref{fig:model}. 
In order to obtain multi-scale feature maps for contextual matching, we carry out feature extraction through three branches, \emph{i.e.}, target LR, reference LR and reference branches. 
Then, the multi-scale feature maps generated from the three branches are fed into the contextual matching module to obtain matched reference features at different scales. 
We then feed these matched features into the multi-scale aggregation blocks (MAB) to guide the upsampling of the target LR at multiple scales, finally obtaining the reconstructed target SR image.

\subsection{Transformer-empowered Feature Extraction}

As mentioned above, the long-range dependencies embedded in the feature maps are essential for efficient and robust contextual matching.  
%
In this regard, we leverage Swin Transformer group (STG) to extract the deep features of each branch; it is capable of extracting deep hierarchical representations with rich long-range dependencies~\cite{liu2021swin} from both target and reference images, which facilitates the proposed network to more efficiently and precisely perform the matching. 
%
As shown in~\cref{fig:model}, the STG consists of multiple residual Swin Transformer blocks (RSTB), each employing multiple Swin Transformer layers (STL) for local attention and cross-window interaction learning. 
The RSTB adopts residual learning to ensure the stability of feature extraction.
A 3$\times$3 convolution layer is used for feature enhancement after RSTBs and STLs. 
%
%
The feature extraction process of RSTB can be expressed as:
\begin{equation}
  \mathcal{F}_{R S T B}=\operatorname{Conv}\left(\mathcal{F}_{S T L}\right)+\mathcal{F}_{\text {in }},
  \label{eq:01}
\end{equation}
where $\mathcal{F}_{STL}$ denotes the features generated from STL, $\operatorname{Conv}$ denotes the 3$\times$3 Conv2D, and $\mathcal{F}_{in}$ denotes the input features of RSTB. As shown in \cref{fig:model}, STL consists of multi-head self-attention blocks and multi-layer perceptions. More details about STL can be found in \cite{liu2021swin}.
In our implementation, we set the number of RSTB and STL as 4 and 6, respectively. 

We extract the multi-scale features from the reference image in a pyramid form, as shown in \cref{fig:model}.
We retain the output of each level of the pyramid and set different stride in Conv2D to ensure that the output of each layer has a different scale, named as $\mathcal{F}_{ref}^i$. 
Afterwards, these features are fed into the multi-scale context matching module for relevant feature mapping. 
The deep features obtained from LR branches are named as $\mathcal{F}_{tar_{-}lr}$ and $\mathcal{F}_{ref_{-}lr}$ respectively. Note that the features obtained in LR branches are on the same scale as the features obtained at the top of the pyramid in the reference branch.

\begin{figure}
  \centering
   \includegraphics[width=\linewidth]{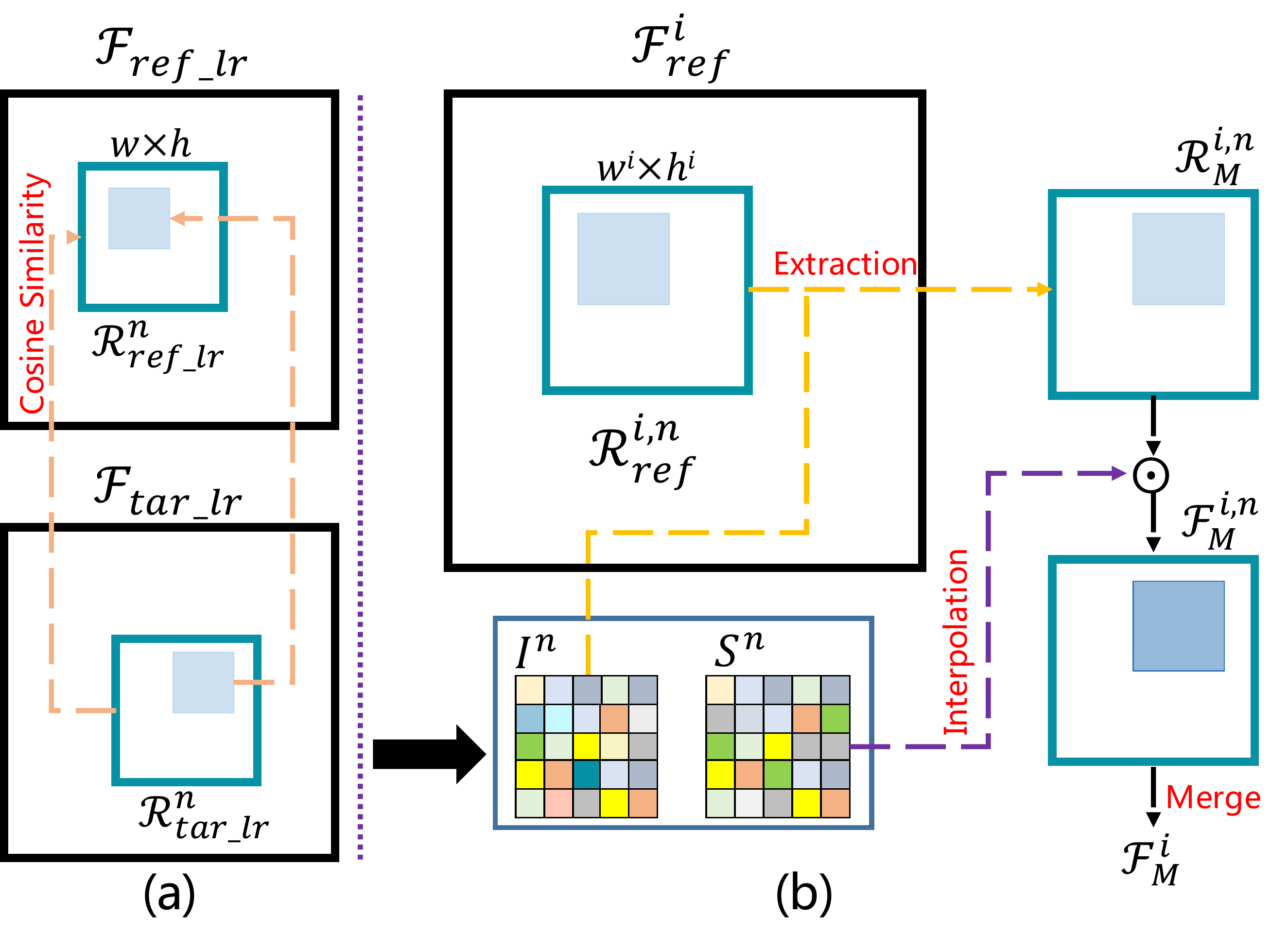}
   \caption{The process of multi-scale context matching, (a) low-scale feature context matching, (b) multi-scale feature mapping.}
   \label{fig:cm}
   \vspace{-5mm}
\end{figure}

\begin{figure*}
  \centering
   \includegraphics[width=\linewidth]{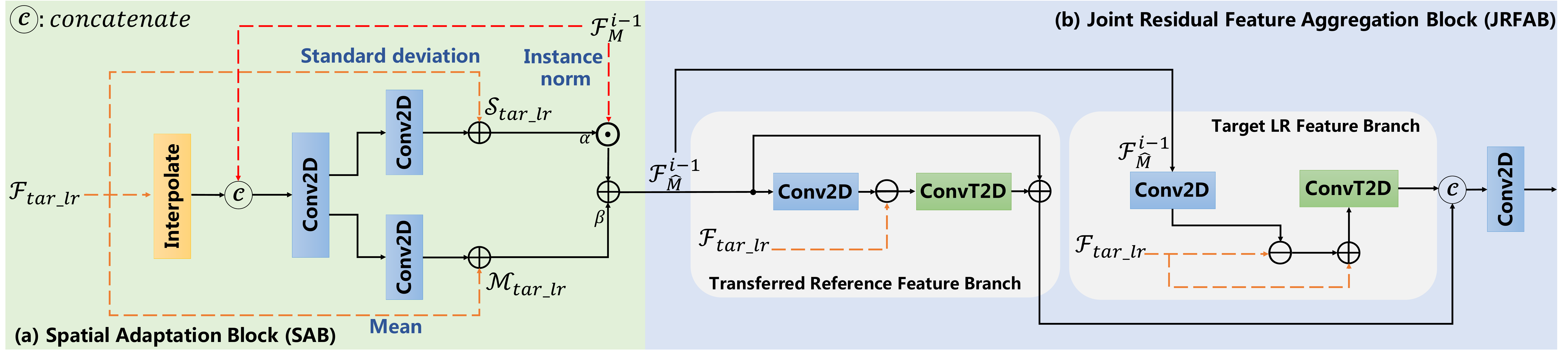}
   \caption{Multi-scale aggregation block, i,e., the fusion strategy of target LR in the upsampling process. \textbf{SAB:} spatial adaptation block, \textbf{JRFAB:} joint residual feature aggregation block. This strategy maximizes the use of the information in the matched reference features. Note that ConvT2D means ConvTranspose2D.}
   \label{fig:UFF}
   \vspace{-5mm}
\end{figure*}
\subsection{Multi-Scale Contextual Matching}
\vspace{-1mm}
Efficient and accurate matching of features is at the core of reference image based SR reconstruction.
%
It is crucial to leverage the details embedded in the SR reference image to make sure the reconstructed SR can contain sufficient anatomical information for clinical applications.
%
%
Traditional matching schemes are incapable of achieving satisfactory results in our task because (1) directly fusing the features extracted from the multi-contrast images may bring redundant yet unnecessary features to the target images and thus reduce the quality of SR images and, (2) owing to the characteristics of medical images, long-range dependencies are quite important to offer more context-aware matching pairs for meaningful SR reconstruction but they are largely neglected in previous schemes.      
%

In this regard, inspired by~\cite{lyu2020multi,lu2021masa}, we perform multi-scale context matching before multi-contrast feature fusion, attempting to obtain the most relevant parts of target and reference features, \emph{i.e.}, $\mathcal{F}_{tar_{-}lr}$ and $\mathcal{F}_{ref_{-}lr}$. 
Then it is mapped to the reference features at different scales, \emph{i.e.}, $\mathcal{F}_{ref}^i$. 
%
%
In addition, thanks to the Transformers equipped in our network, we can implicitly harness the long-range dependencies embedded in the extracted features to enhance the matching quality. 
As shown in \cref{fig:cm}, our context matching can be divided into two steps: 1) context matching of low-scale features $\mathcal{F}_{tar_{-}lr}$ and $\mathcal{F}_{ref_{-}lr}$ to obtain index and similarity maps, and 2) mapping them into multi-scale features $\mathcal{F}_{ref}^i$. The details are elaborated as follows.

\textbf{1) Low-scale feature context matching.}
To reduce the computational cost of the network, we compute the similarity maps in the target and reference features on the low-scale features. 
We first expand $\mathcal{F}_{t a r_{-} l r}$ into $N$ non-overlapping blocks to get $\mathcal{R}_{t a r_{-} l r}^n$ ($1 \leq n \leq N$); the patch size is $w \times h$ (where UF=4, $w$=$h$=13).
Then, we take each $\mathcal{R}_{tar_{-}lr}^n$ patch center region to calculate the cosine similarity value to find the center region with the greatest similarity to $\mathcal{F}_{ref_{-}lr}$, and get $\mathcal{R}_{ref_{-}lr}^n$ patch.
We crop $\mathcal{F}_{ref}^i$ with this center region to obtain multi-scale similar patches with size of $w^i \times h^i$, named as $\mathcal{R}_{ref}^{i,n}$. 
Thus, for each $\mathcal{R}_{tar_{-}lr}^n$ patch, there is a corresponding most relevant $\mathcal{R}_{ref_{-}lr}^n$ and $\mathcal{R}_{ref}^{i,n}$ patch.
Note that as all feature maps are generated from STGs, the long-range dependencies embedded in them will implicitly affect the matching, enhancing the similarity values among patches with similar anatomical structures but located separately.  
Next, we perform region matching on $\mathcal{R}_{t a r_{-} l r}^n$ and $\mathcal{R}_{ref_{-} l r}^n$ to get index maps $\mathcal{I}^n$ and similarity maps $\mathcal{S}^n$. 
For example,
we first compute the similarity value between $z$-th region of $\mathcal{R}_{tar_{-}lr}^n$ and $g$-th region of $\mathcal{R}_{ref_{-}lr}^n$ to get $s_{z,g}^n$.
Then, we compute the $z$-th elements of the index map $\mathcal{I}^n$ and similarity map $\mathcal{S}^n$, as follows:
\begin{equation}
\mathcal{I}_{z}^{n}=\underset{g}{\operatorname{argmax}} \; s_{z, g}^{n} \quad 
\mathcal{S}_{z}^{n}=\underset{g}\max \; s_{z, g}^{n}.
\label{eq:02}
\end{equation}
Please refer to \cite{lu2021masa} for details on how to calculate similarity values.

\textbf{2) Multi-scale feature mapping.}
After getting the index and similarity maps, we have to map them to the $\mathcal{R}_{ref}^{i,n}$ patch at different scales to ensure that the reference features at multiple scales all contain the most similar features to the target LR, as shown in \cref{fig:cm} (b).
Specifically, according to $\mathcal{I}^n$, we extract related regions $\mathcal{R}_M^{i,n}$ from $\mathcal{R}_{ref}^{i,n}$ patch.
Then, we multiply $\mathcal{R}_M^{i,n}$ with the corresponding similarity map $\mathcal{S}^n$, and get the weighted features block $\mathcal{F}_M^{i,n}$. 
Note that $i$ represents the reference features of different scale sizes. 
As the similarity map $\mathcal{S}^n$ is obtained on the LR scale, when $i > 1$, interpolation is required for $\mathcal{S}^n$. 
The above process is formulated as:
\begin{equation}
\mathcal{F}_{M}^{i, n}=multiply\left(\mathcal{R}_{M}^{i, n}, u p\left(\mathcal{S}^{n}\right)\right),
\label{eq:03}
\end{equation}
where $multiply$ and $up$ denote multiplication and bilinear interpolation. Finally, we merge $N$ patches, and obtain multi-scale matched reference features \emph{i.e.}, $\mathcal{F}_M^i$.

\begin{figure*}[t]
	\centering
	\captionsetup[subfloat]{labelformat=empty}
	\subfloat[\scriptsize Tar LR/HR]{
	\begin{minipage}[b]{0.1\textwidth}
    	\includegraphics[scale=0.205]{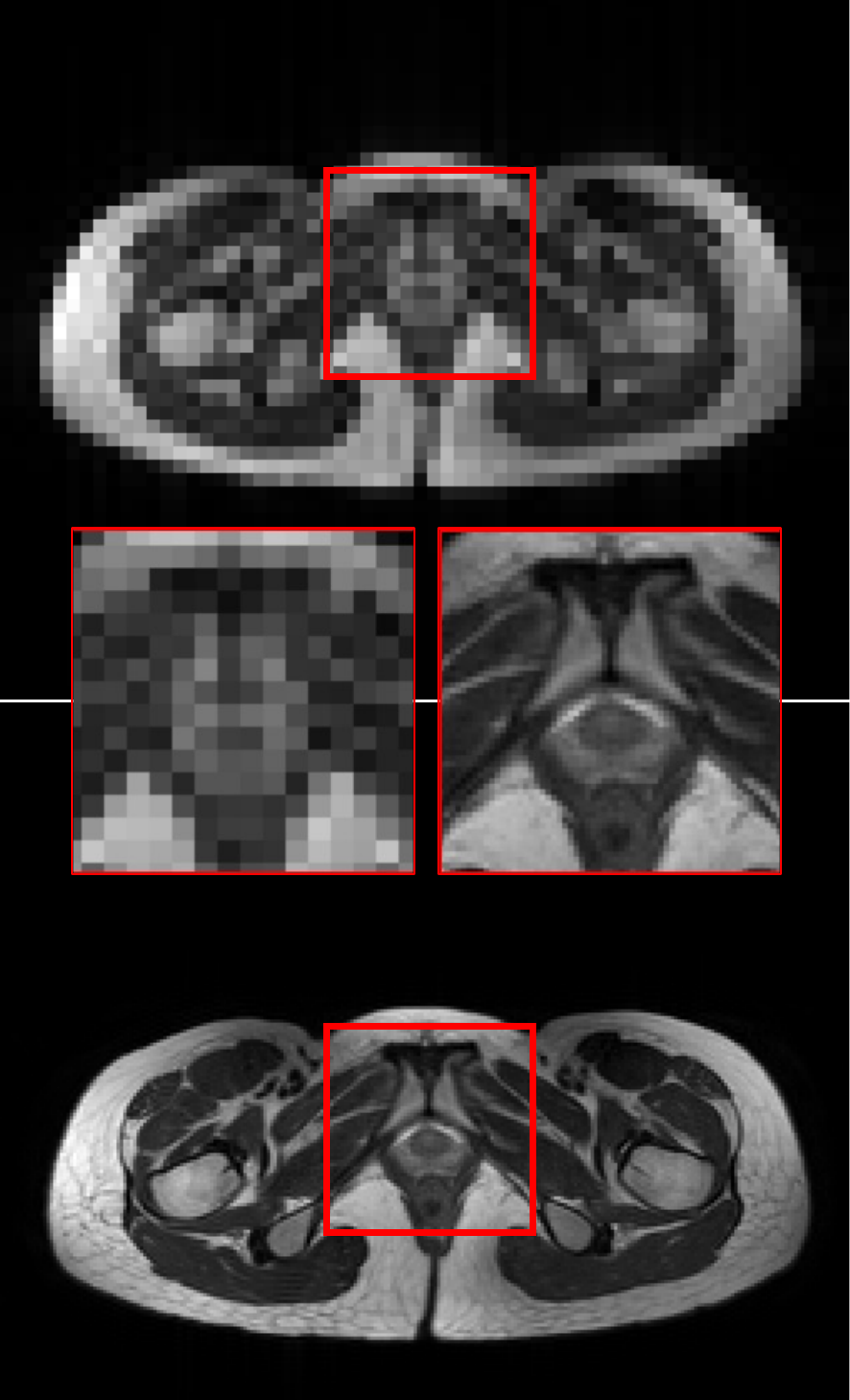}
	\end{minipage}
}
\hfill
	\subfloat[\scriptsize Bicubic]{
	\begin{minipage}[b]{0.1\textwidth}
		\includegraphics[scale=0.205]{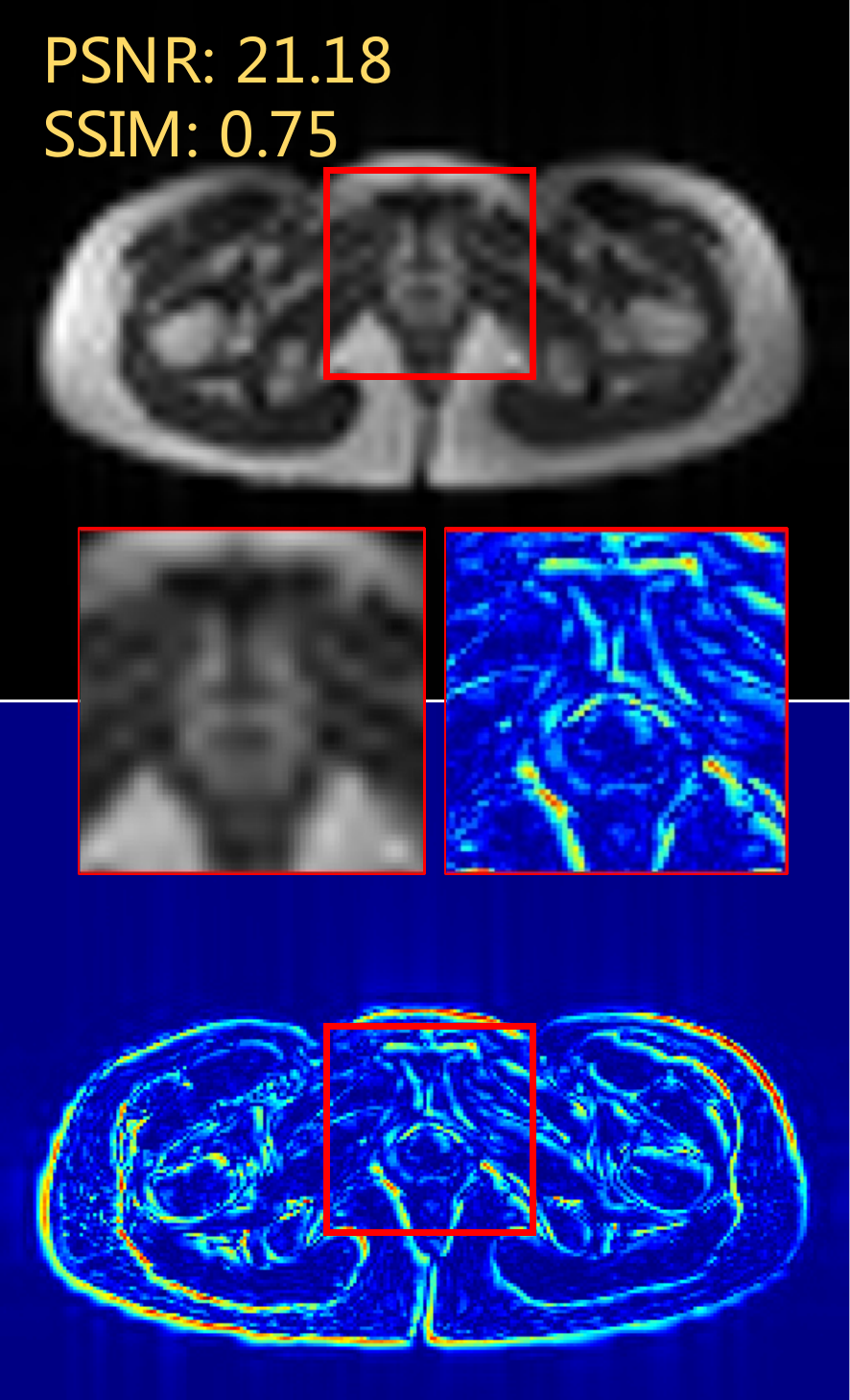} 
	\end{minipage}
}
\hfill
	\subfloat[\scriptsize EDSR \cite{ lim2017enhanced }]{
	\begin{minipage}[b]{0.1\textwidth}
		\includegraphics[scale=0.205]{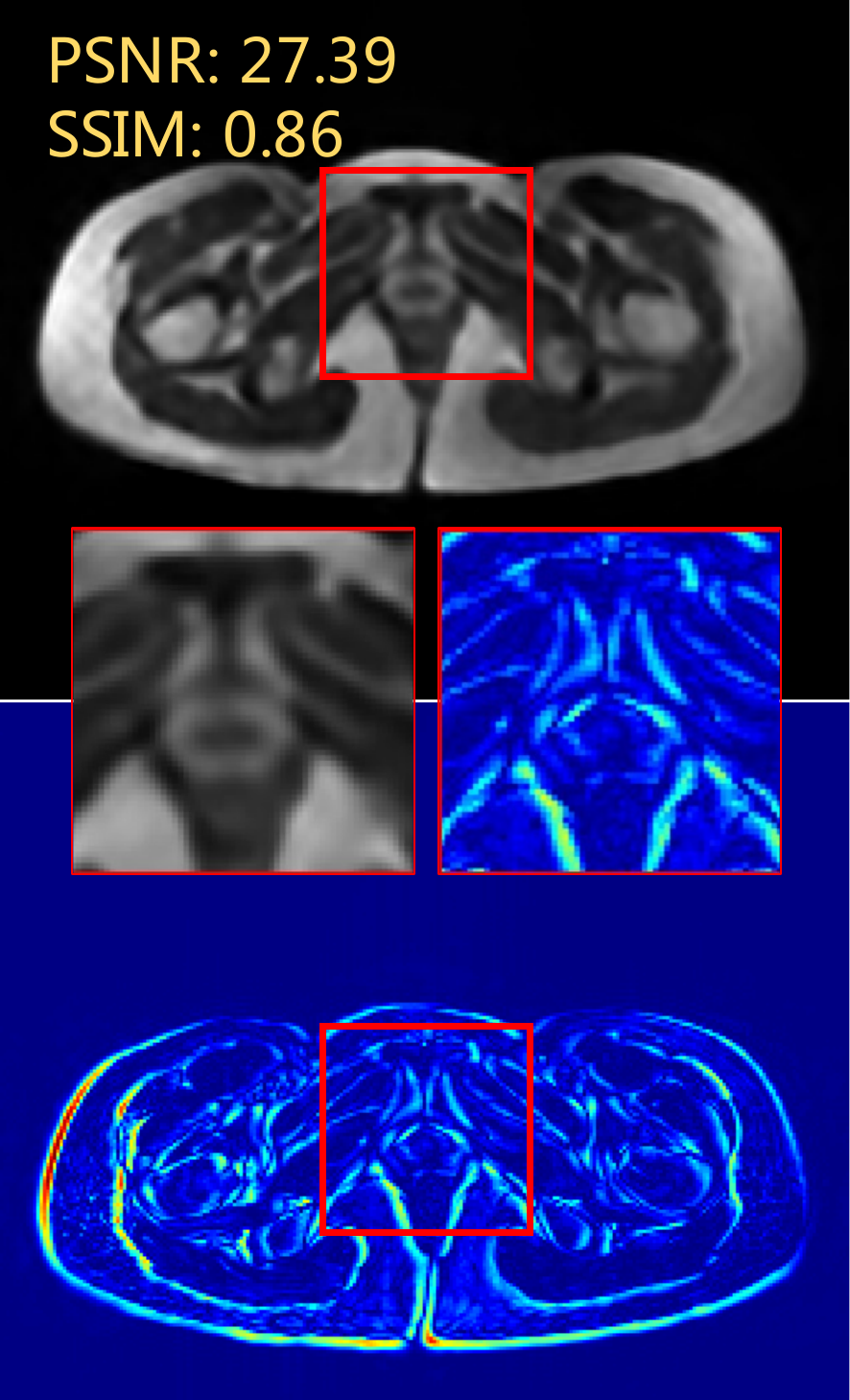} 
	\end{minipage}
}
\hfill
    \subfloat[\scriptsize MCSR \cite{ lyu2020multi }]{
	\begin{minipage}[b]{0.1\textwidth}
		\includegraphics[scale=0.205]{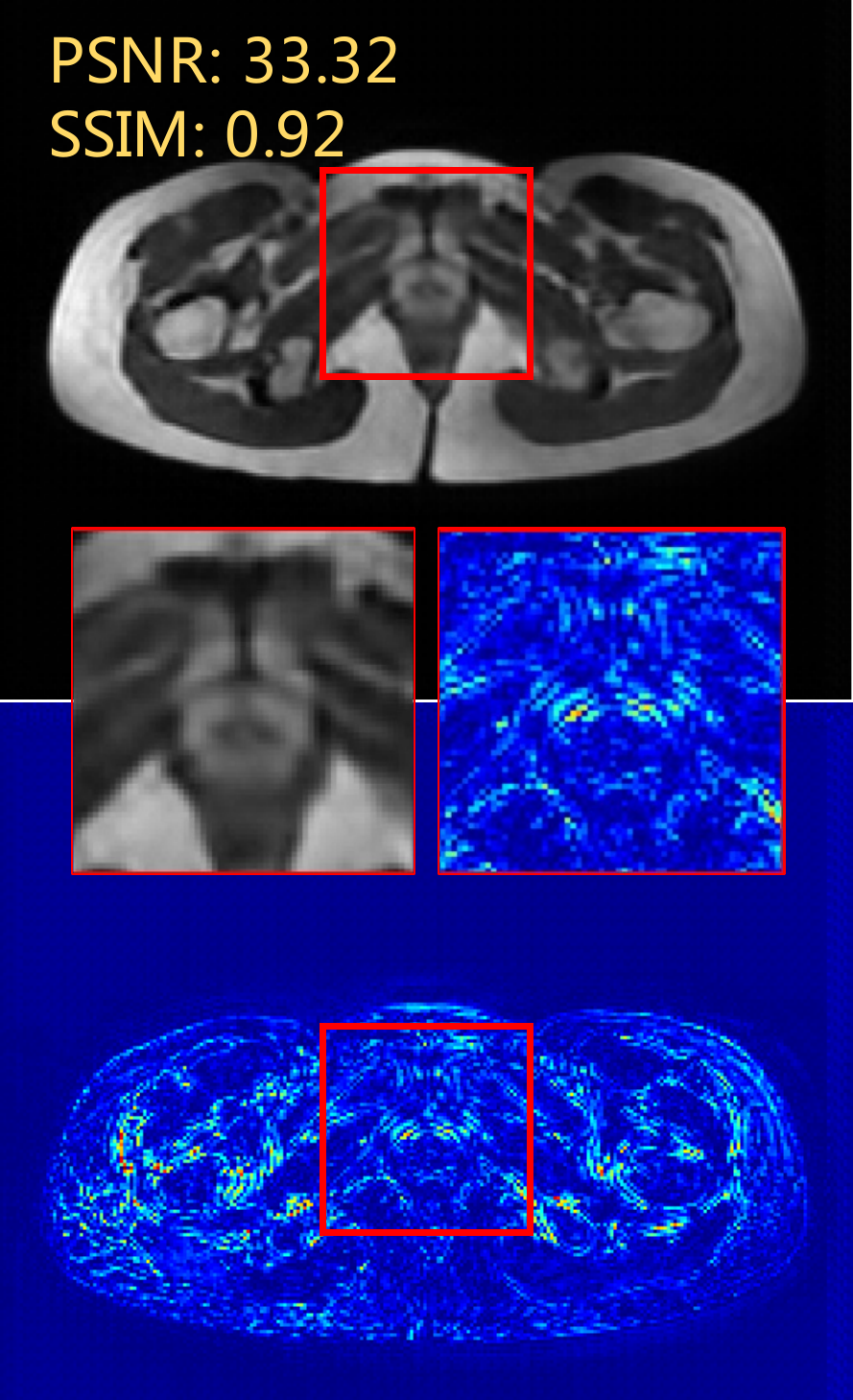} 
	\end{minipage}
}
\hfill
    \subfloat[\scriptsize MINet \cite{ feng2021multi }]{
	\begin{minipage}[b]{0.1\textwidth}
		\includegraphics[scale=0.205]{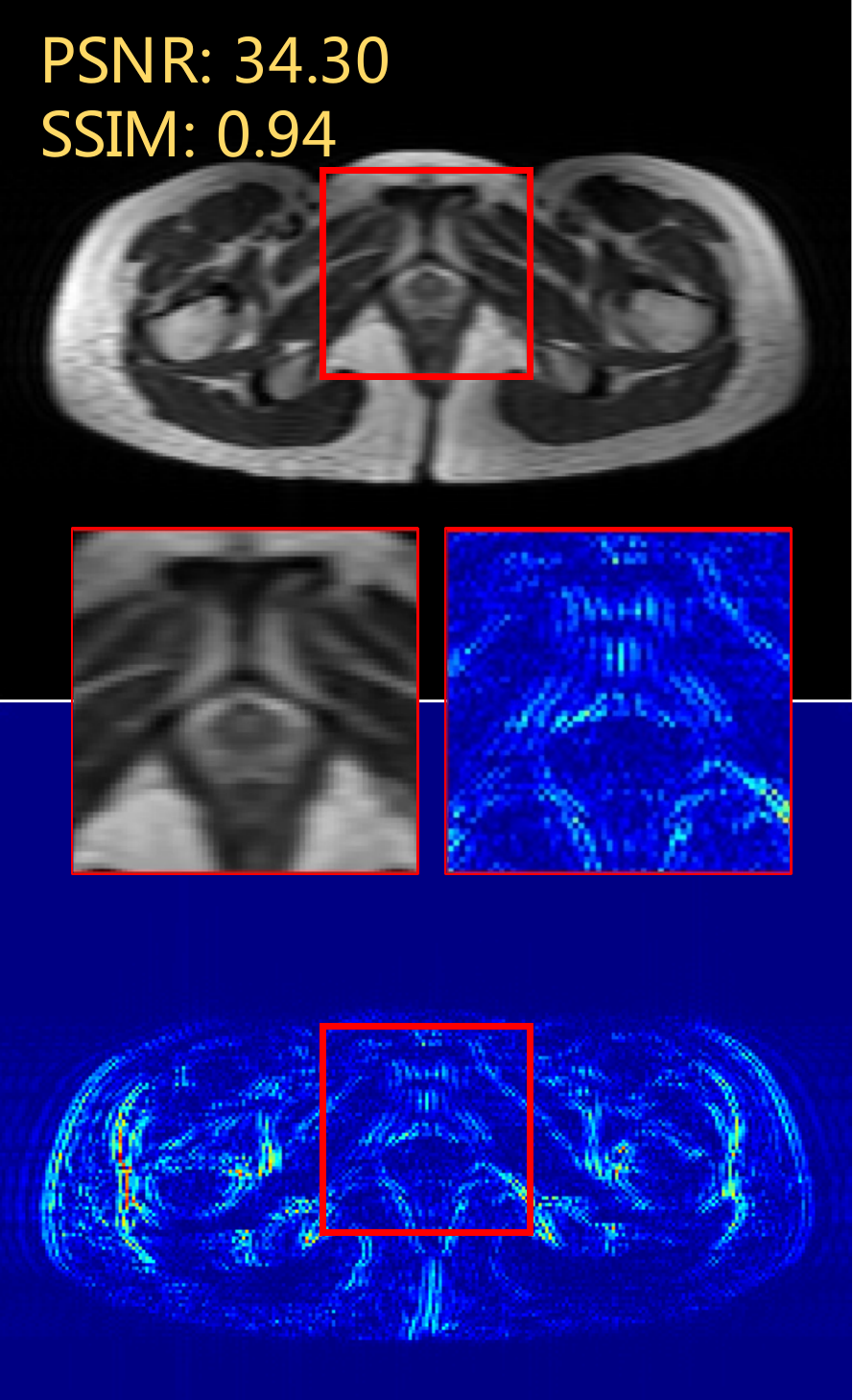} 
	\end{minipage}
}
\hfill
    \subfloat[\scriptsize MASA \cite{ lu2021masa }]{
	\begin{minipage}[b]{0.1\textwidth}
		\includegraphics[scale=0.205]{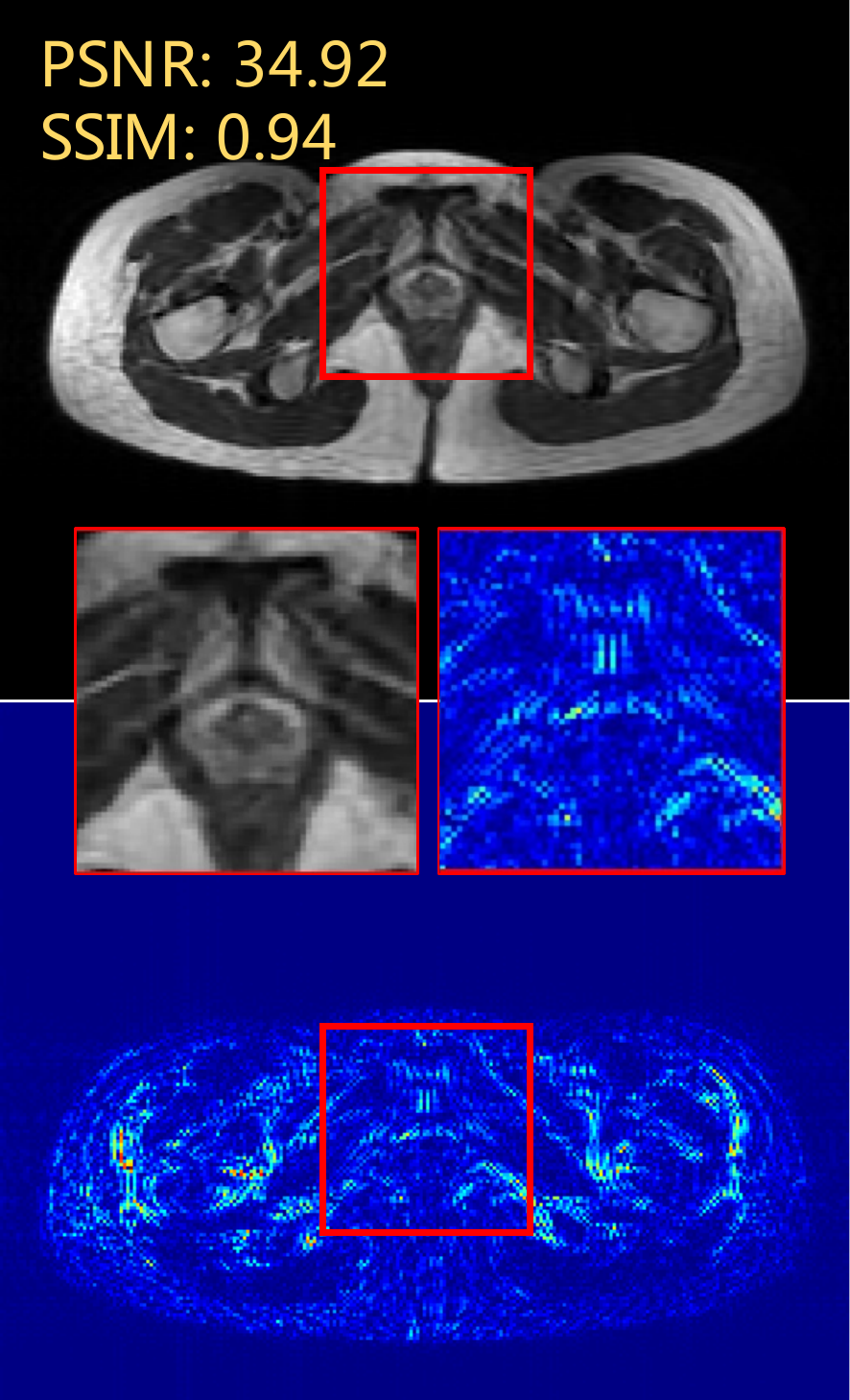} 
	\end{minipage}
}
\hfill
    \subfloat[\scriptsize SwinIR \cite{liang2021swinir}]{
	\begin{minipage}[b]{0.1\textwidth}
		\includegraphics[scale=0.205]{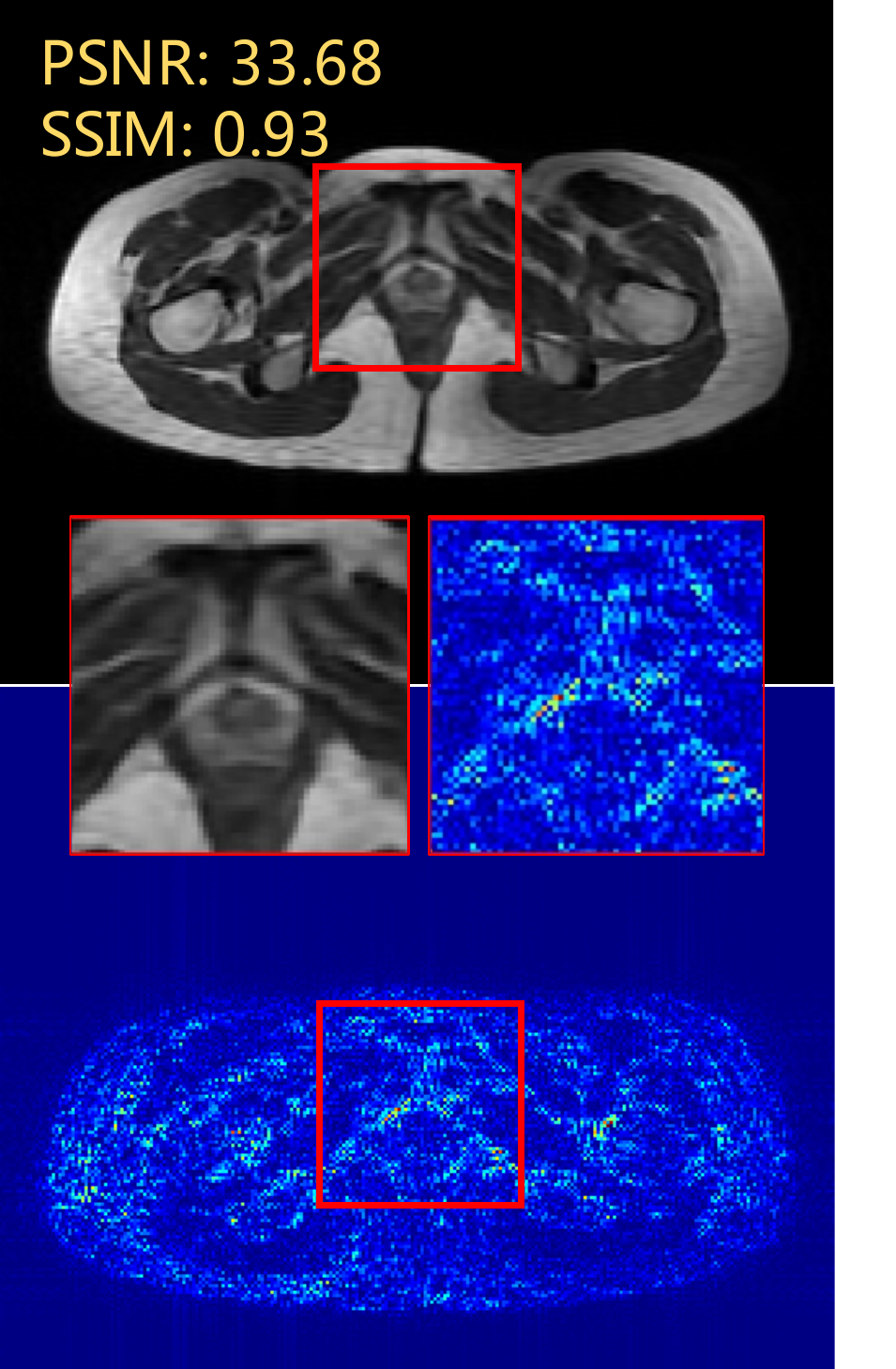} 
	\end{minipage}
}
\hfill
    \subfloat[\scriptsize Restormer \cite{zamir2021restormer}]{
	\begin{minipage}[b]{0.1\textwidth}
		\includegraphics[scale=0.205]{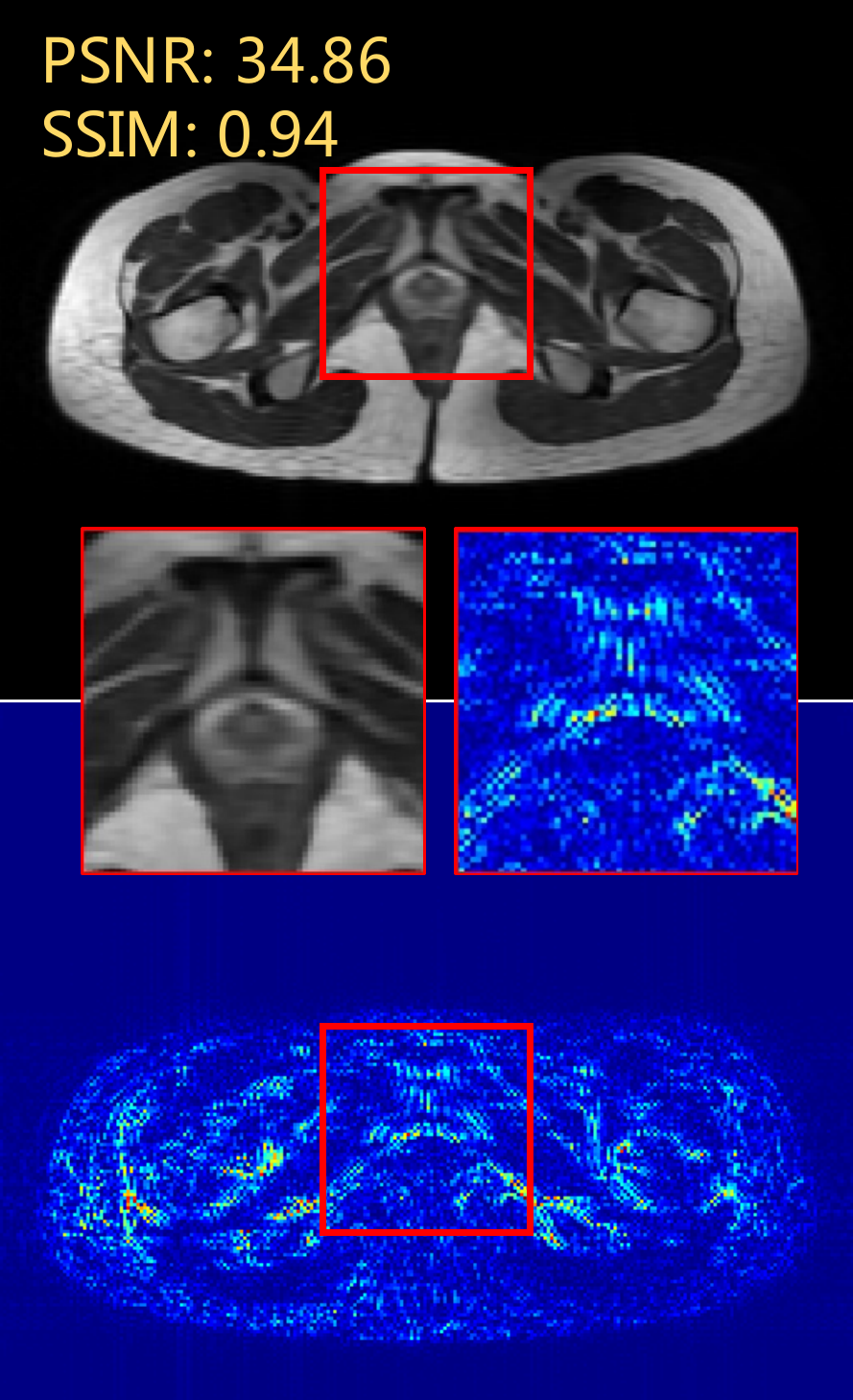} 
	\end{minipage}
}
\hfill
    \subfloat[\scriptsize McMRSR]{
	\begin{minipage}[b]{0.1\textwidth}
		\includegraphics[scale=0.205]{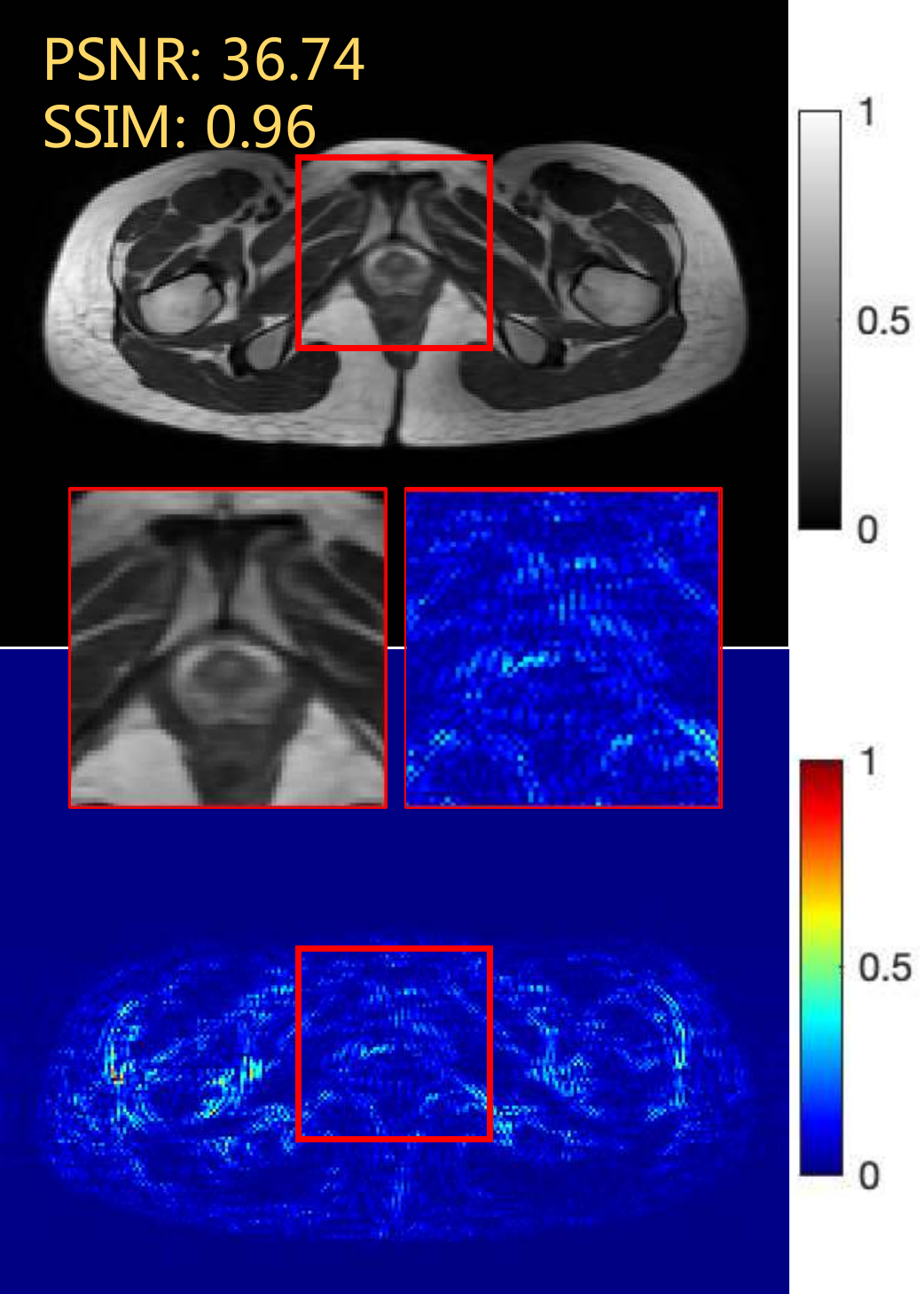} 
	\end{minipage}
}
\caption{Qualitative results of different SR reconstruction methods on pelvic dataset with UF=4. The reconstructed images and the corresponding error maps are provided. The McMRSR recovers fine anatomical structures, as shown in the inset image.}
\vspace{-3mm}
\label{fig:pelvic}
\end{figure*}

\subsection{Multi-Scale Feature Aggregation}
After obtaining multi-scale matched reference features, how to fuse them into the target LR features is an important yet challenging step. 
For the low-scale targeted LR features, fusing the matched reference features at different scales in the upsampling stage can make full use of the matched similar information and recover the details in the image to the maximum. Therefore, inspired by~\cite{lu2021masa}, we design $MAB_{i}$ (the number corresponds to $Scale_{i}$) to help target LR aggregate multi-scale matched reference features, \emph{i.e.}, $\mathcal{F}_M^i$. As shown in \cref{fig:model}, the low-scale target LR features aggregate the features matched at the top of the pyramid, and then sequentially aggregate the reference features at different scales. This approach ensures that the matched features are fully utilized for the target LR features at each scale during upsampling. As shown in \cref{fig:UFF}, this block consists of a spatial adaptation block (SAB) and a joint residual feature aggregation block (JRFAB). 

\textbf{Spatial Adaptation Block.}
We know that multi-contrast MR images have different colors and brightnesses for different contrasts, although they mirror the same anatomical structures.
The previous multi-contrast MRI SR methods \cite{feng2021multi,lyu2020multi} simply combine the reference and target features together and then perform the next convolution operation, which is not the optimal choice. To enhance the consistency of the matched reference features with the target LR feature distribution, inspired by \cite{ park2019semantic }, we use SAB to remap the distribution of matched reference features onto the distribution of traget LR features. 

As shown in \cref{fig:UFF} (a), the target LR features are upsampled by 2$\times$, and then connected with the matched reference features $\mathcal{F}_M^{i-1}$. 
We use stride of 1, 3$\times$3 Conv2D to get the two parameters $\alpha$ and $\beta$. 
We then figure out the standard deviation and mean of the unsampled target LR features, and calculate $\mathcal{S}_{t a r_{-} l r}$ and $\mathcal{M}_{t a r_{-} l r}$ to update $\alpha$ and $\beta$.
Next, we perform instance normalization \cite{ ulyanov2016instance } on $\mathcal{F}_M^{i-1}$, and the operation is performed with $\alpha$ and $\beta$ to obtain the transferred reference features $\mathcal{F}_{\hat{M}}^{i-1}$ as:
\begin{equation}
\mathcal{F}_{\hat{M}}^{i-1}=multiply(\mathcal{F}_{M}^{i-1} ,\alpha)+\beta.
\label{eq:04}
\end{equation}

\textbf{Joint Residual Feature Aggregation Block.}
After SAB, we obtain the transferred reference features. In order to make the multi-scale features more fully aggregated, we consider further refining the high-frequency details in the transferred reference and target features so as to ensure that the aggregated features assimilate more anatomical details. 
We adopt the JRFAB to divide the aggregation process into two branches, \emph{i.e.}, transferred reference branch and target LR branch, as shown in \cref{fig:UFF} (b). Transferred reference branch is used to enhance the high-frequency details in $\mathcal{F}_{\hat{M}}^{i-1}$, which can be formulated as:
\begin{equation}
\tilde{\mathcal{F}}_{\hat{M}}^{i-1}=\mathcal{F}_{\hat{M}}^{i-1}+\operatorname{ConvT} \left(\operatorname{Conv}\left(\mathcal{F}_{\hat{M}}^{i-1}\right)-\mathcal{F}_{\text {tar }_{-} l r}\right),
\label{eq:05}
\end{equation}
where $\operatorname{Conv}$ denotes 3$\times$3 Conv2D with stride of 2, and $\operatorname{ConvT}$ denotes 3$\times$3 ConvTranspose2D with stride 2. 
Similarly, the refinement of high-frequency information in the target LR features can be expressed as:
\begin{equation}
\tilde{\mathcal{F}}_{\text {tar}_{-} l r }=\operatorname{ConvT}\left(\mathcal{F}_{\text {tar}_{-} l r }+\left(\mathcal{F}_{\text {tar }_{-} l r}-\operatorname{Conv}\left(\mathcal{F}_{\hat{M}}^{i-1}\right)\right)\right).
\label{eq:06}
\end{equation}
Finally, we concatenate the output of the two branches and get the output of $MAB_{i-1}$ after 3$\times$3 Conv2D with stride 1. Note that when $\mathcal{F}_M^i$ has the same scale as $\mathcal{F}_{tar_{-}lr}$, upsampling of LR features is not required in SAB, and Conv2D is used in JRFAB instead of ConvT2D.

\begin{figure*} [t]
	\centering
	\captionsetup[subfloat]{labelformat=empty}
	\subfloat[\scriptsize Tar LR/HR]{
	\begin{minipage}[b]{0.1\textwidth}
    	\includegraphics[scale=0.205]{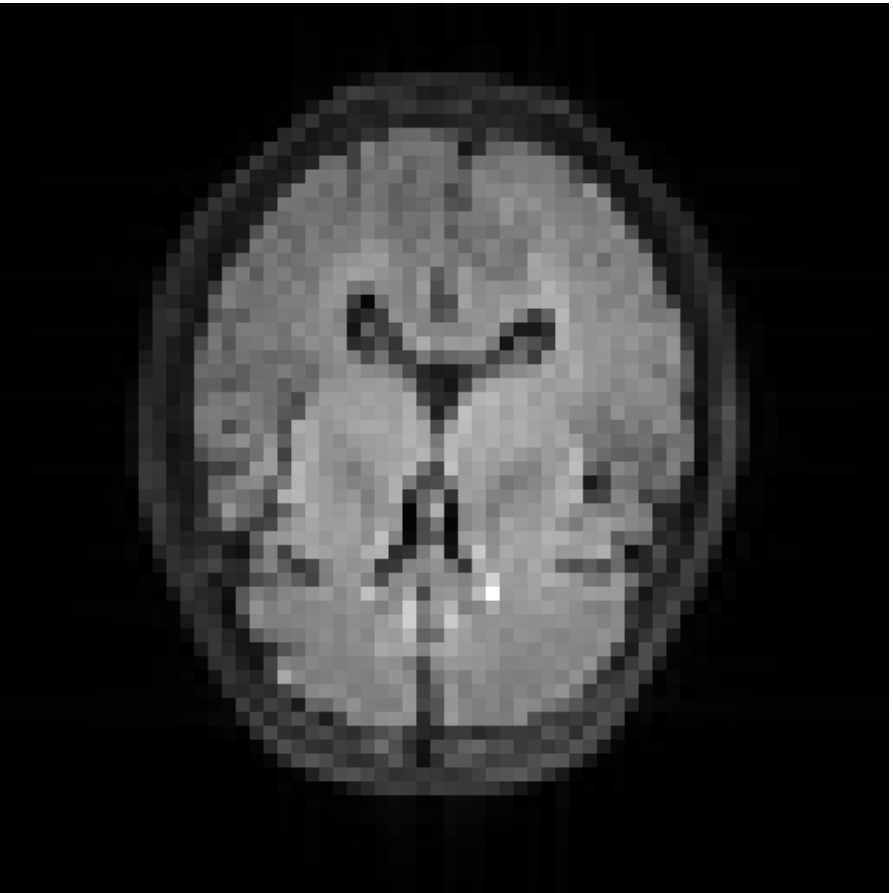}
	    \\
		\includegraphics[scale=0.205]{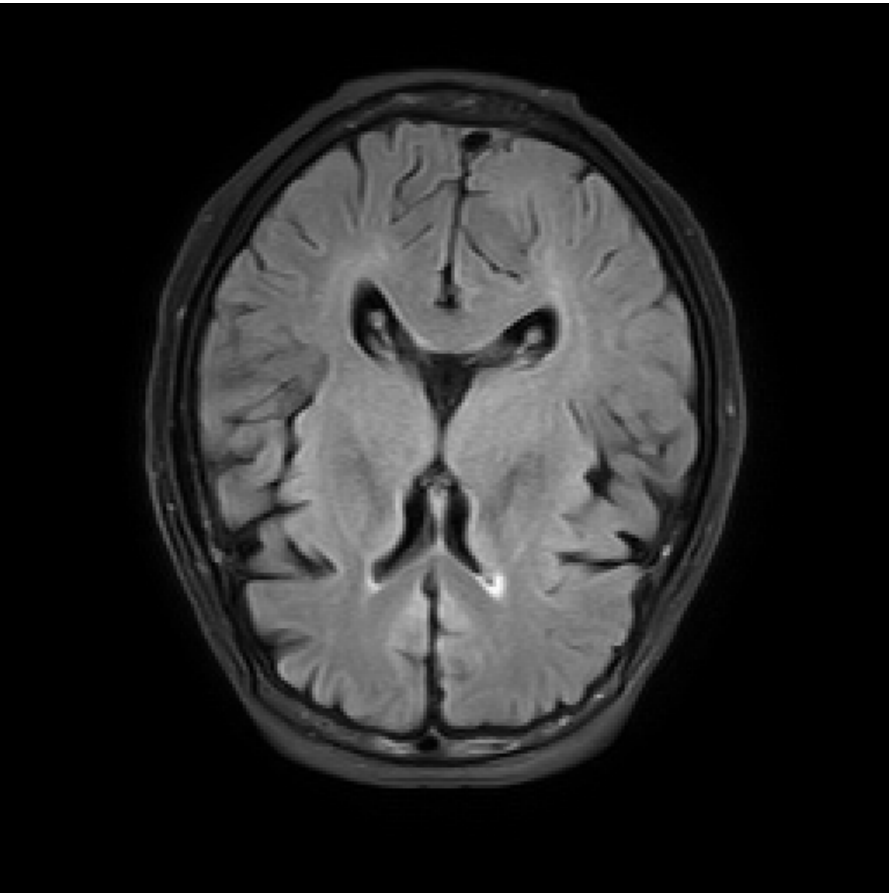}
		\\
		\includegraphics[scale=0.205]{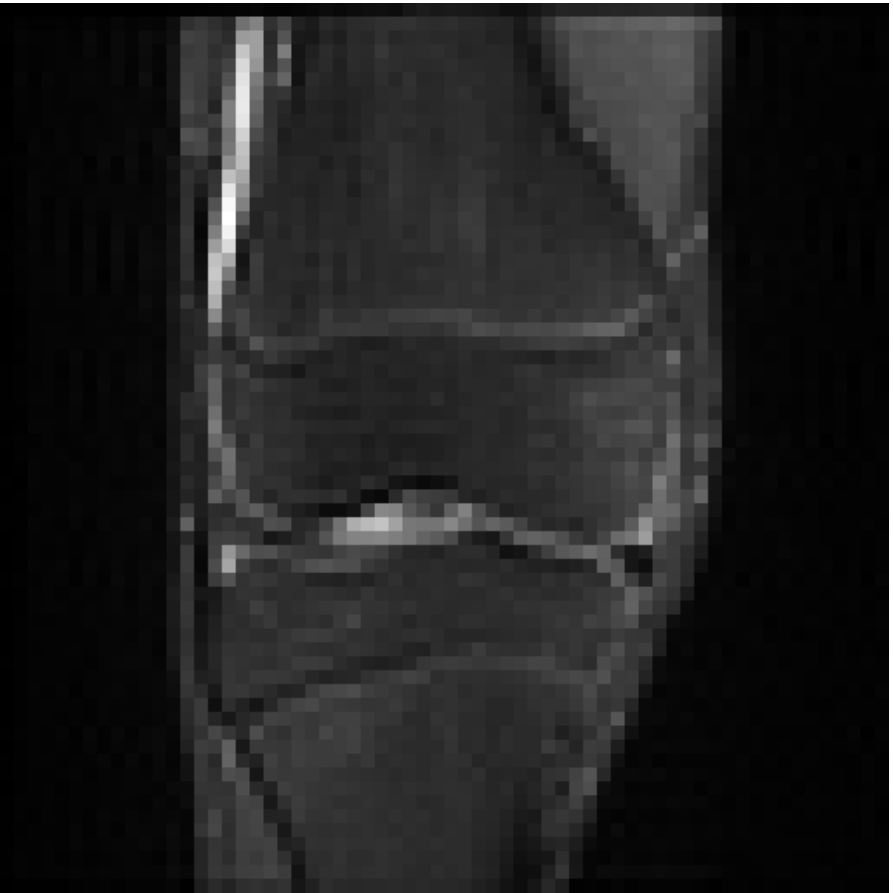}
		\\
		\includegraphics[scale=0.205]{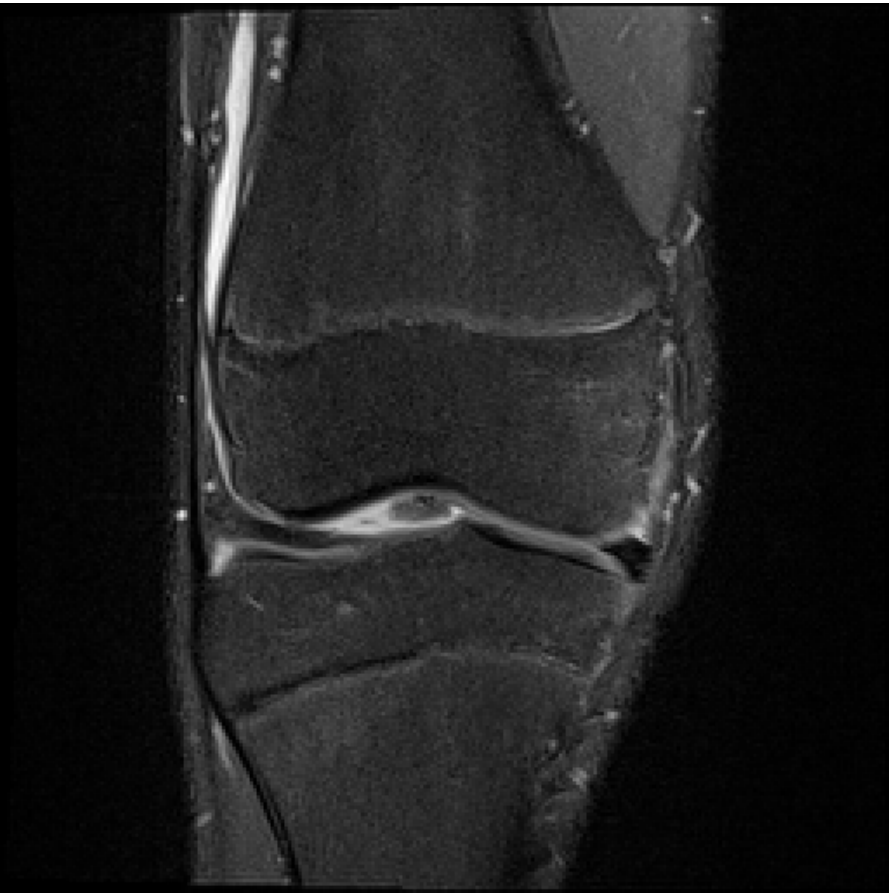}
	\end{minipage}
}
\hfill
	\subfloat[\scriptsize Bicubic]{
	\begin{minipage}[b]{0.1\textwidth}
		\includegraphics[scale=0.205]{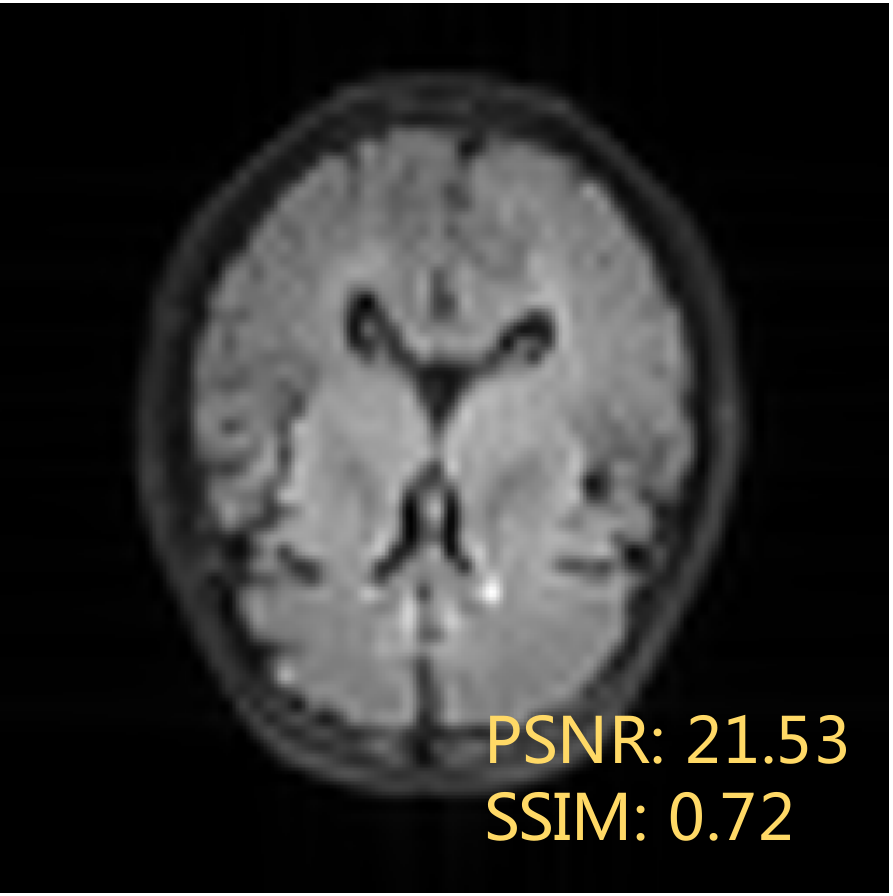} 
		\\
    	\includegraphics[scale=0.205]{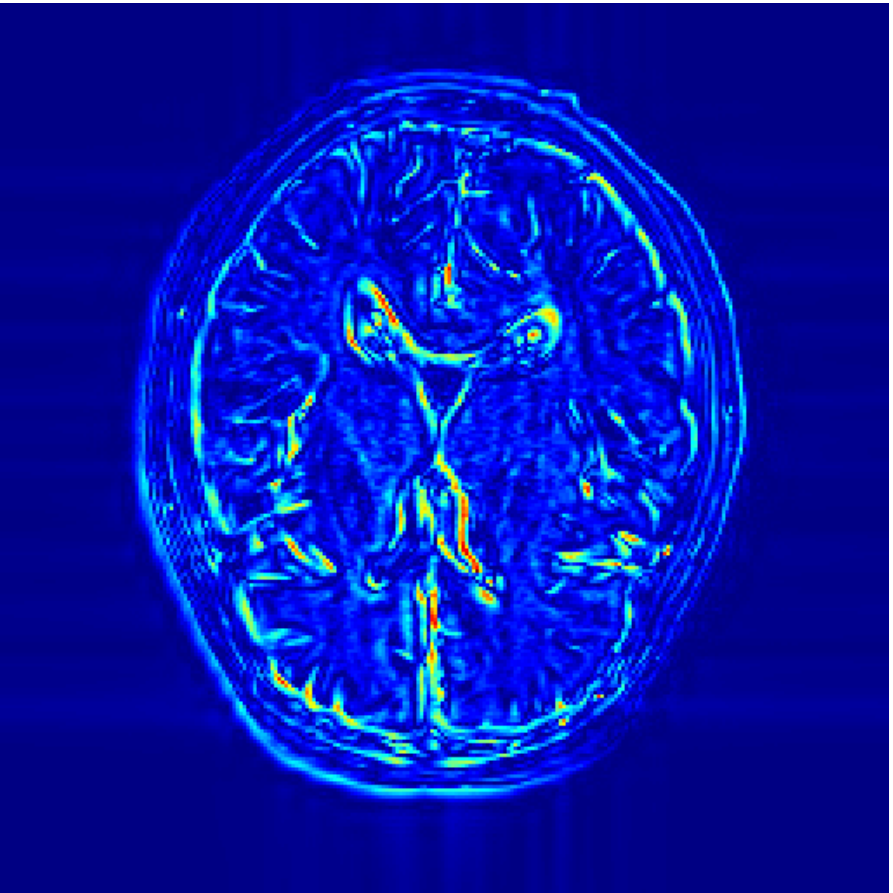}
    	\\
		\includegraphics[scale=0.205]{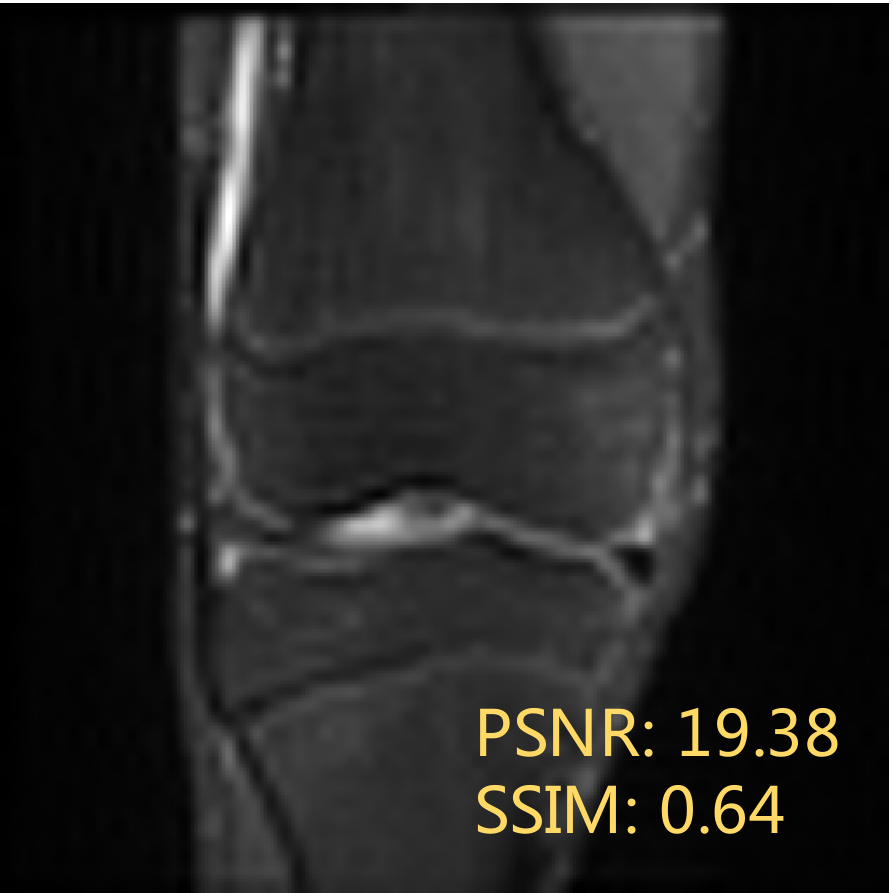}
		\\
		\includegraphics[scale=0.205]{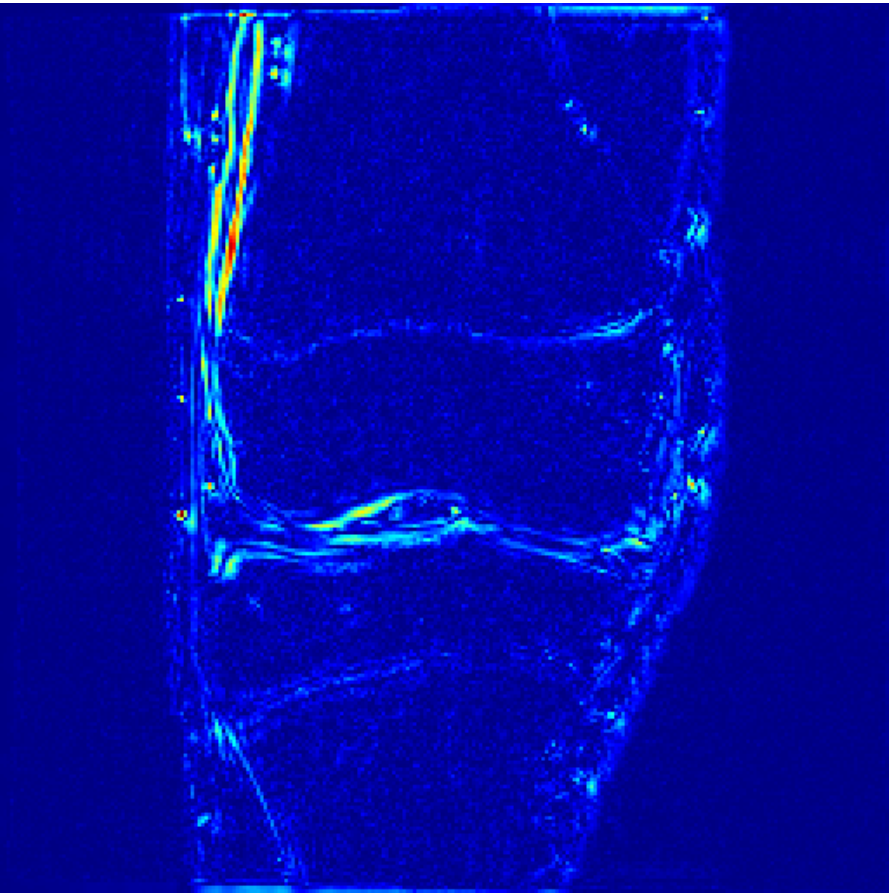}
	\end{minipage}
}
\hfill
	\subfloat[\scriptsize EDSR \cite{ lim2017enhanced }]{
	\begin{minipage}[b]{0.1\textwidth}
		\includegraphics[scale=0.205]{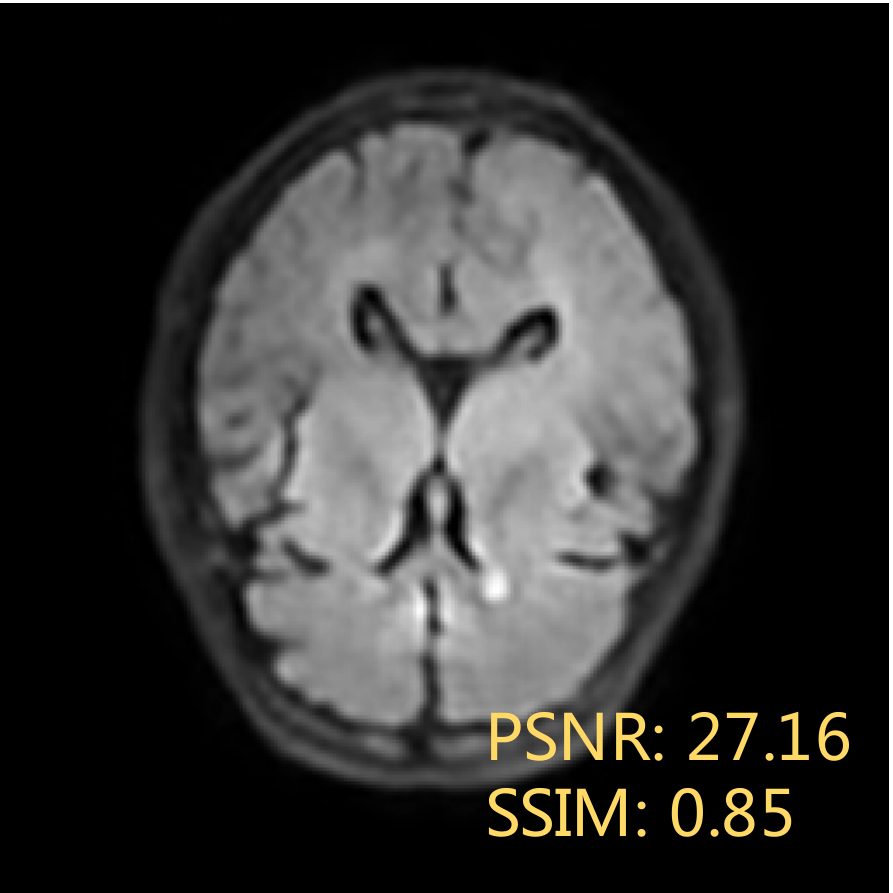} 
		\\
		\includegraphics[scale=0.205]{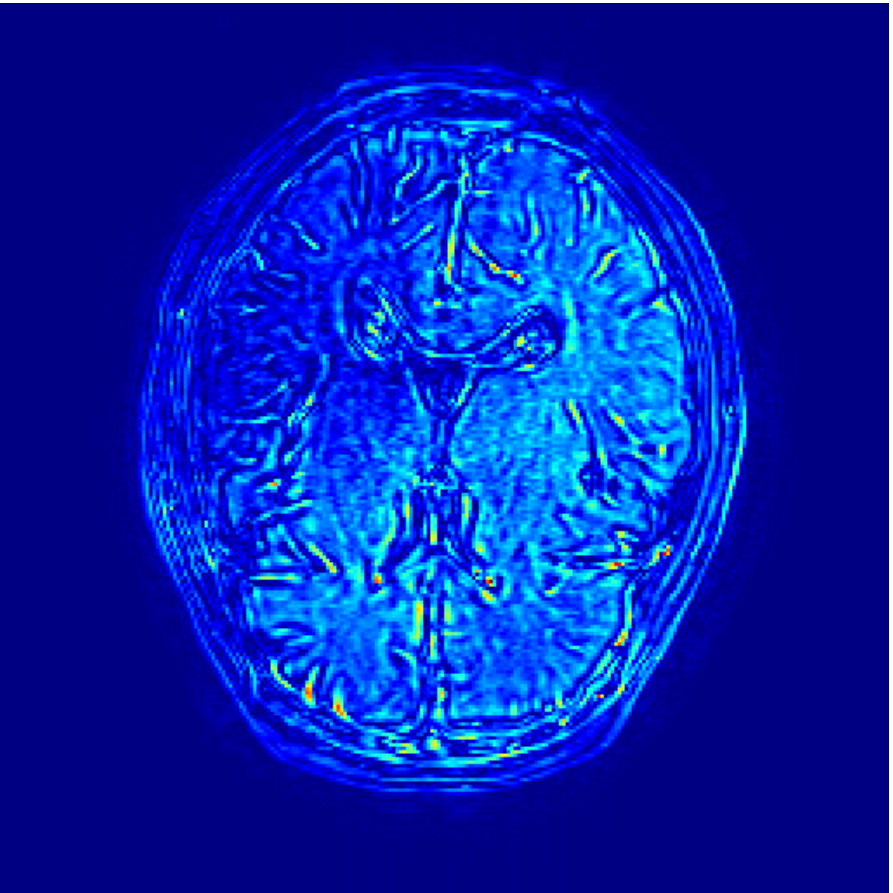}
		\\
		\includegraphics[scale=0.205]{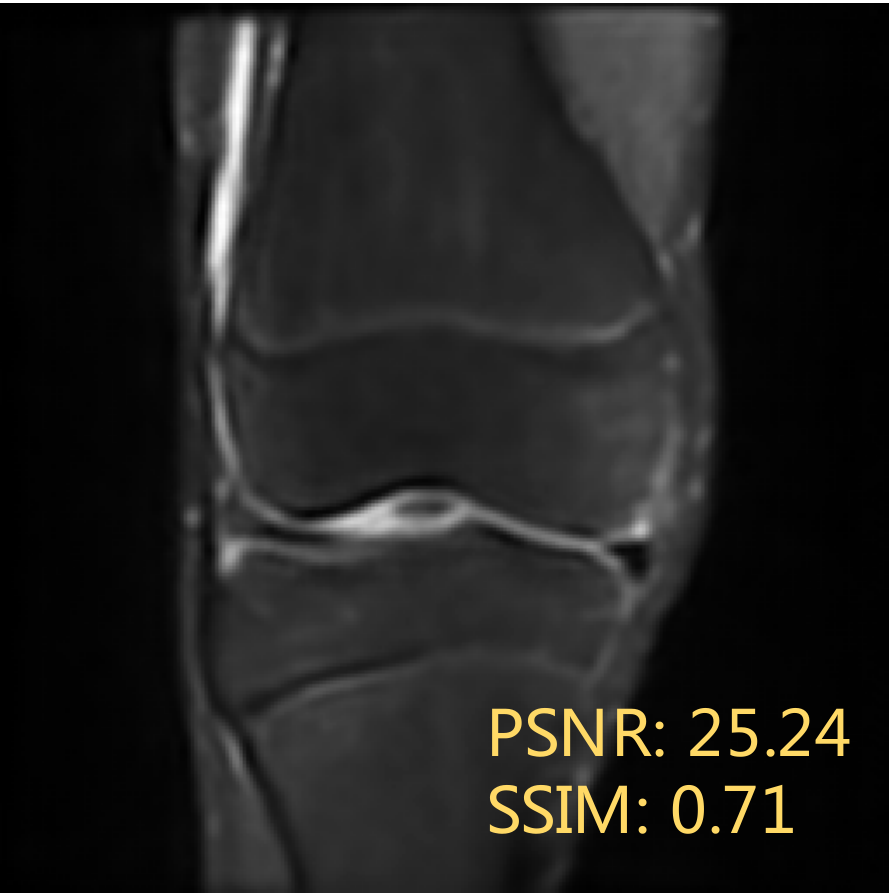}
		\\
		\includegraphics[scale=0.205]{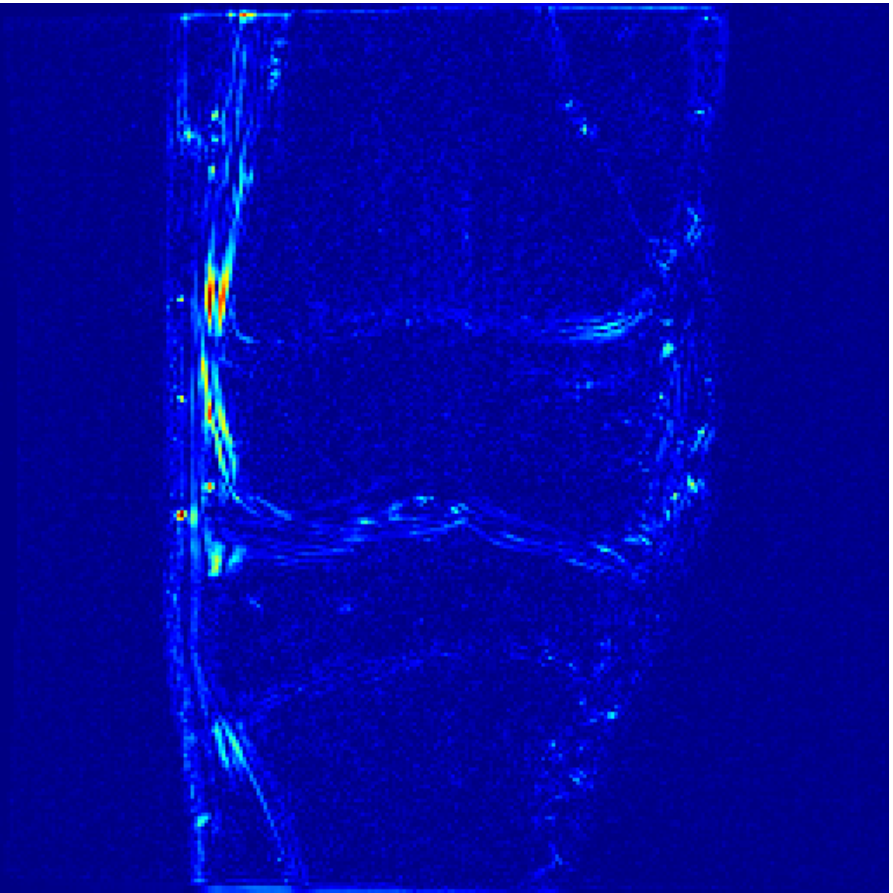}
	\end{minipage}
}
\hfill
    \subfloat[\scriptsize MCSR \cite{ lyu2020multi }]{
	\begin{minipage}[b]{0.1\textwidth}
		\includegraphics[scale=0.205]{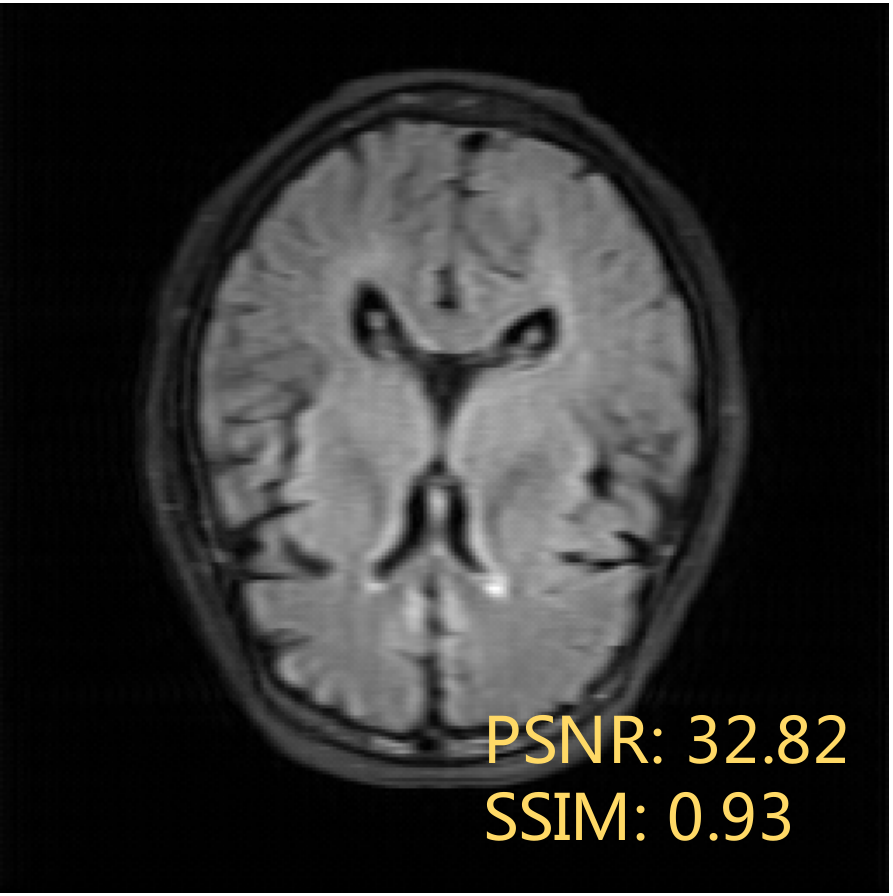} 
		\\
		\includegraphics[scale=0.205]{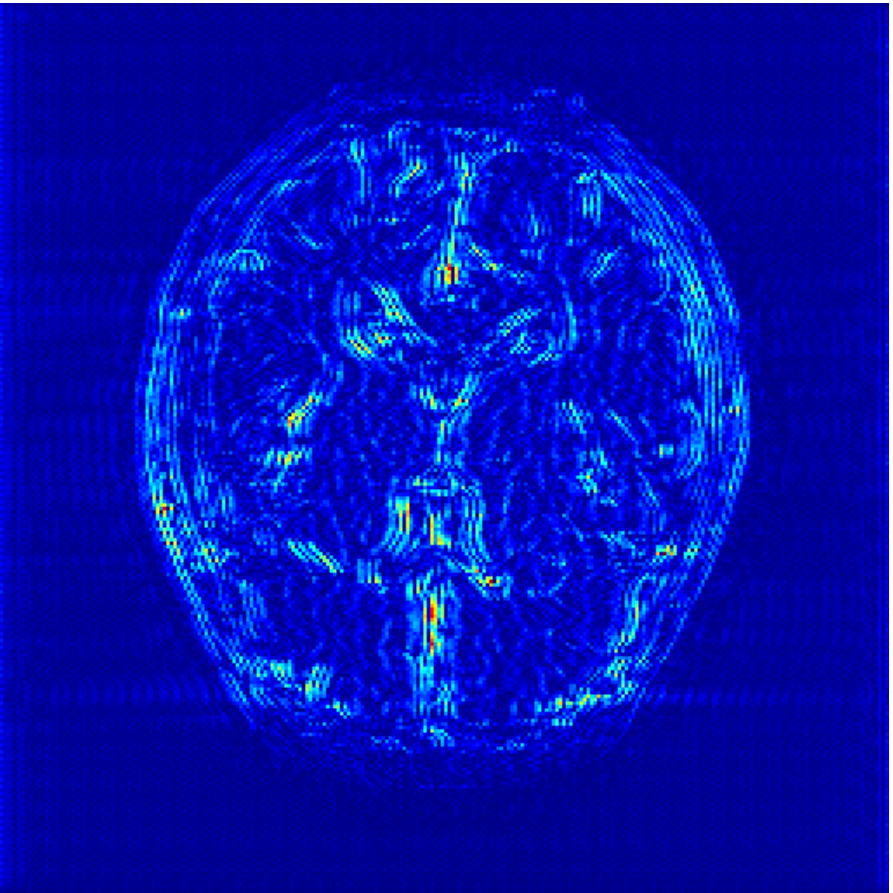}
		\\
		\includegraphics[scale=0.205]{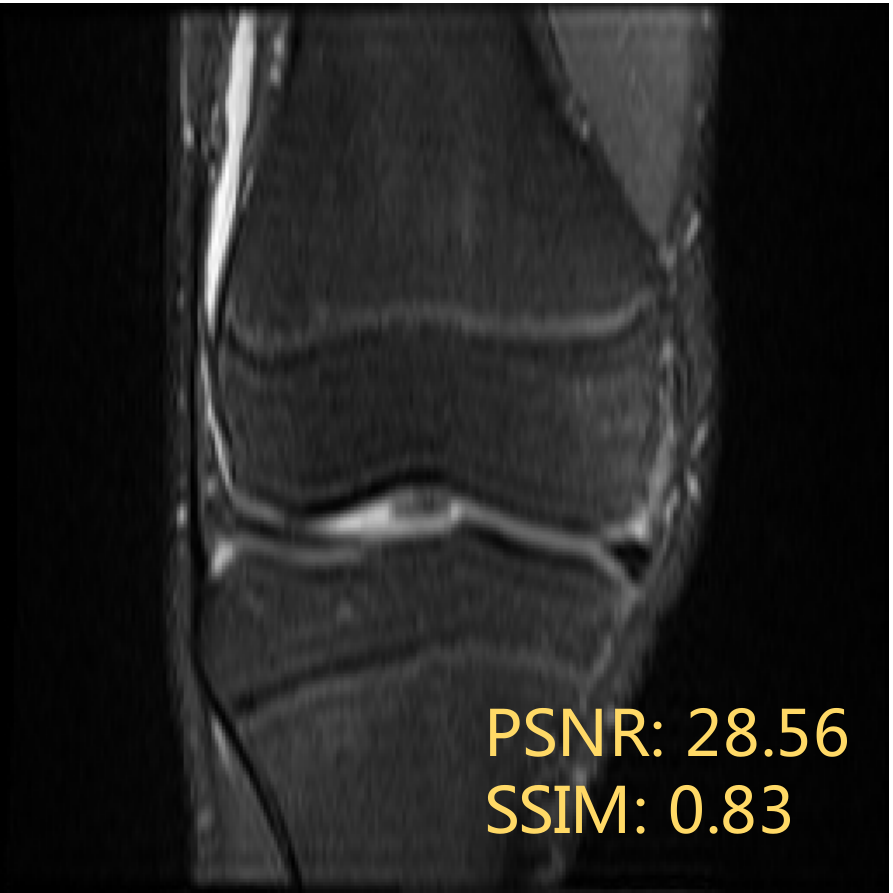}
		\\
		\includegraphics[scale=0.205]{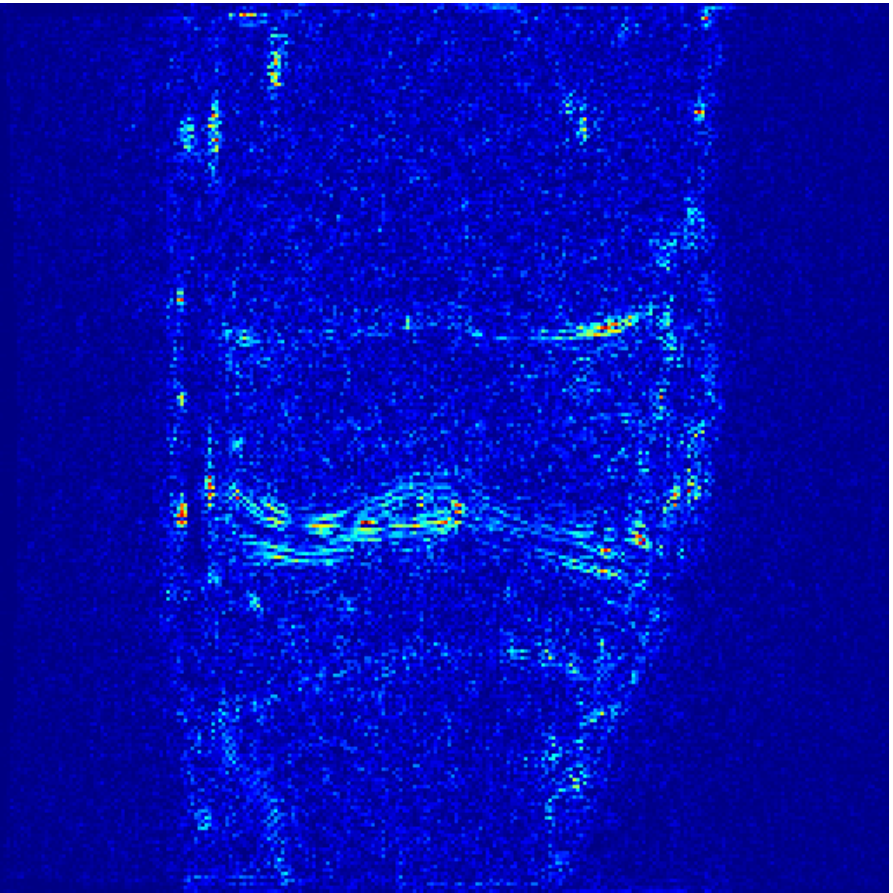}
	\end{minipage}
}
\hfill
    \subfloat[\scriptsize MINet \cite{ feng2021multi }]{
	\begin{minipage}[b]{0.1\textwidth}
		\includegraphics[scale=0.205]{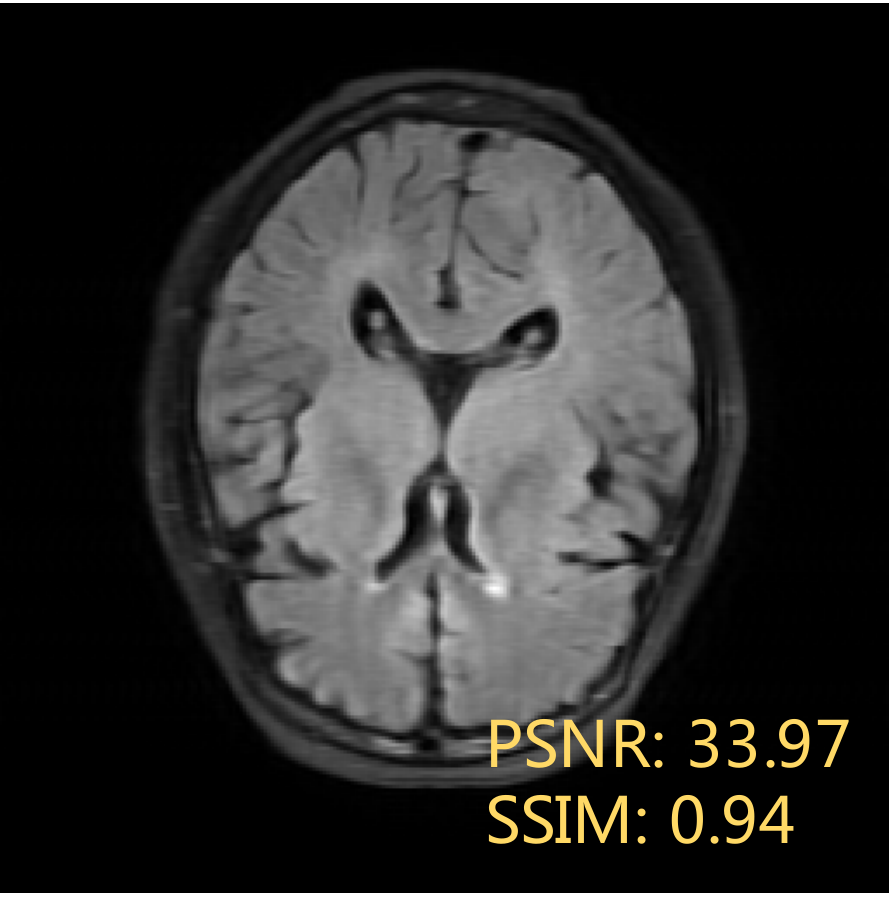} 
		\\
		\includegraphics[scale=0.205]{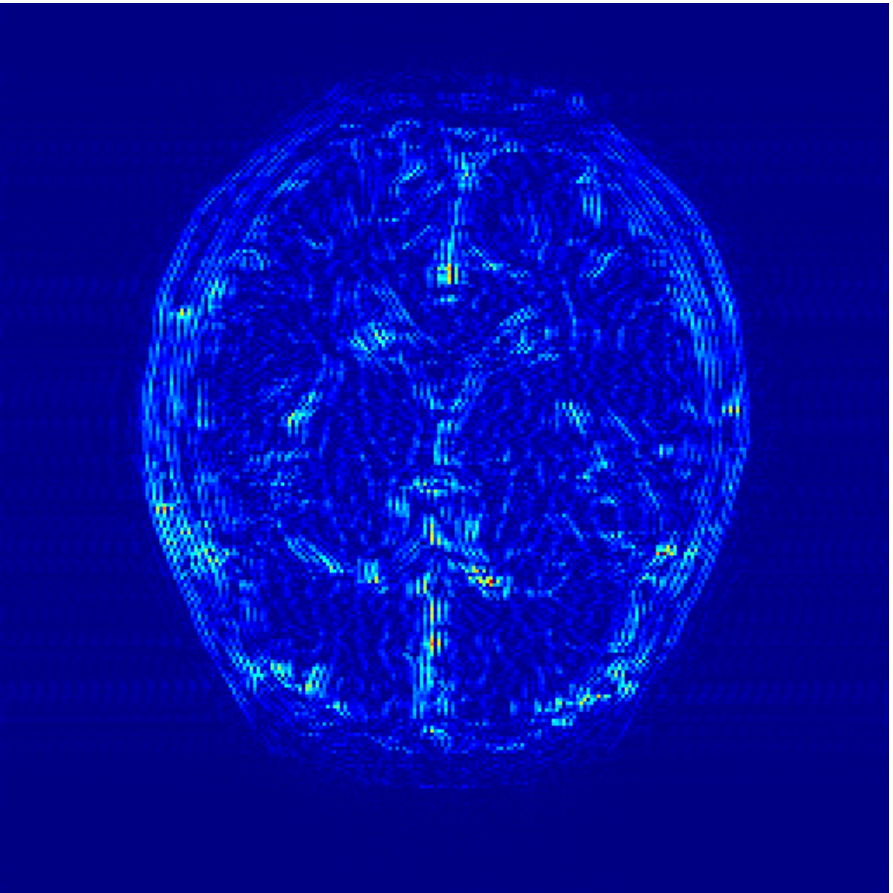}
		\\
		\includegraphics[scale=0.205]{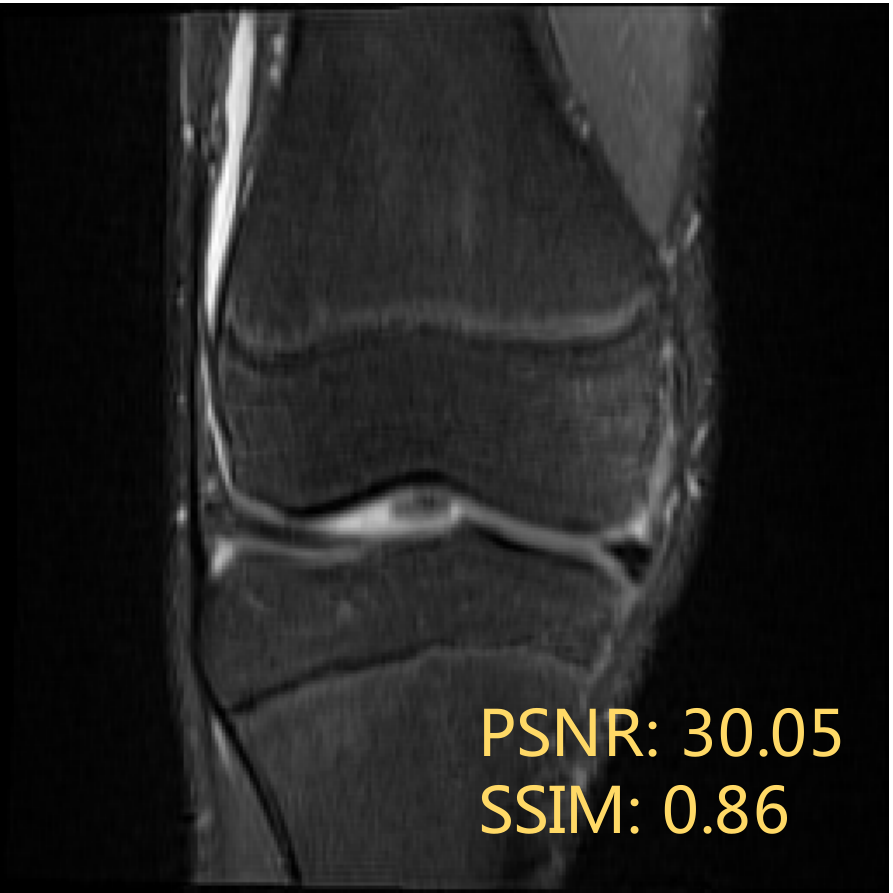}
		\\
		\includegraphics[scale=0.205]{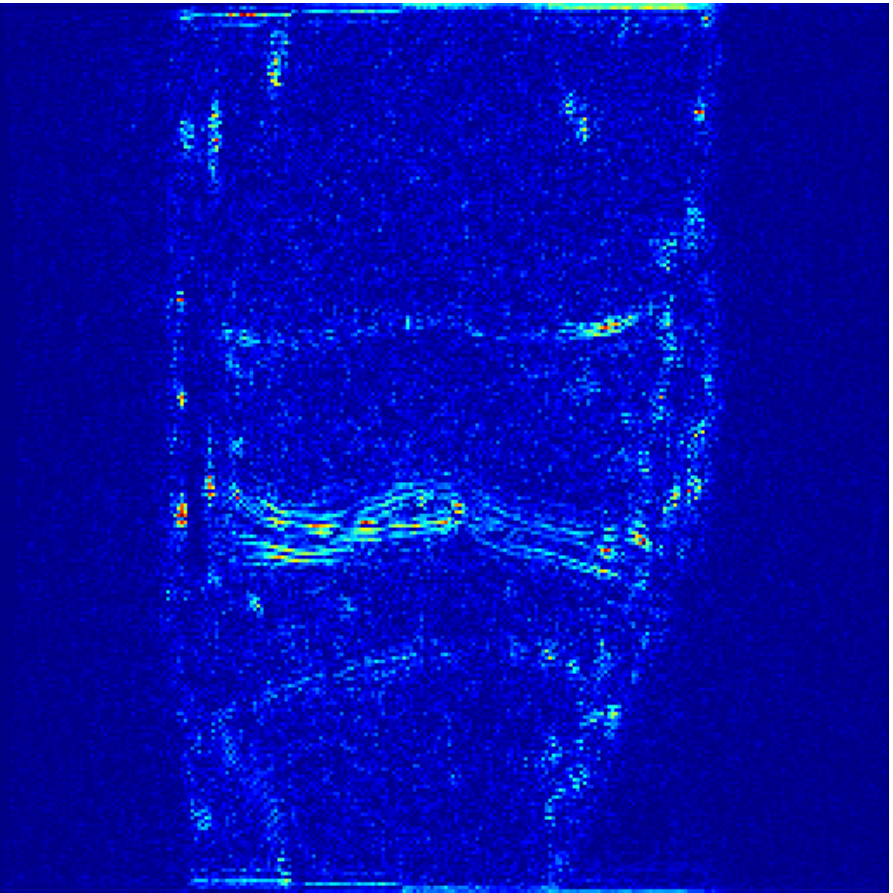}
	\end{minipage}
}
\hfill
    \subfloat[\scriptsize MASA \cite{ lu2021masa }]{
	\begin{minipage}[b]{0.1\textwidth}
		\includegraphics[scale=0.205]{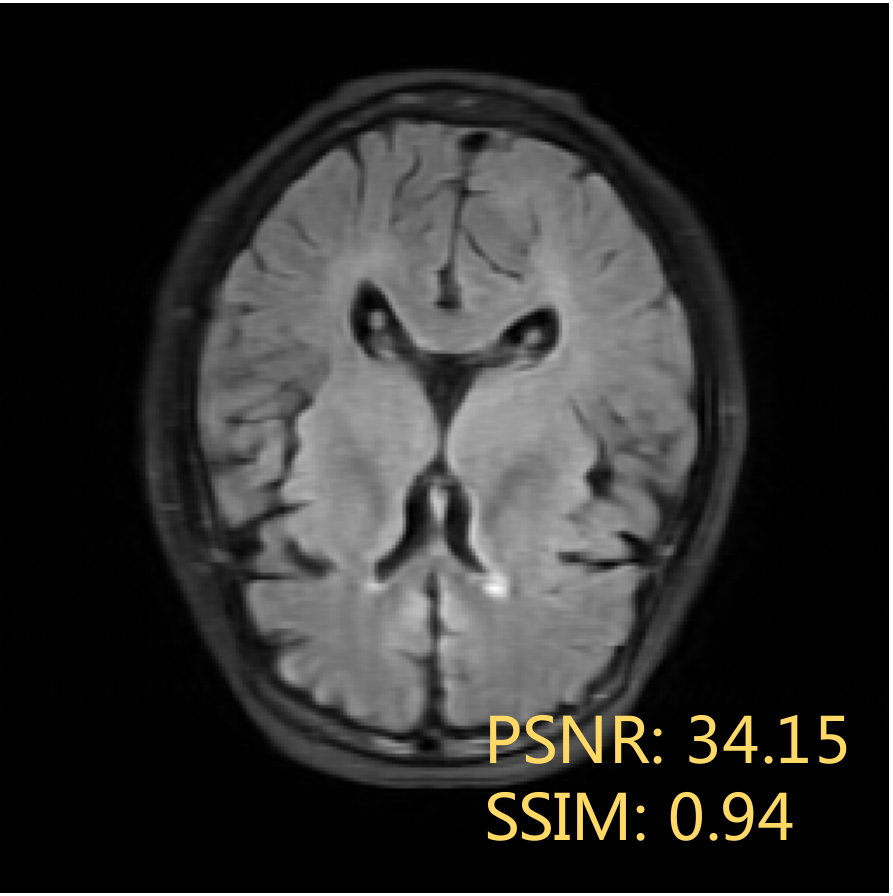} 
		\\
		\includegraphics[scale=0.205]{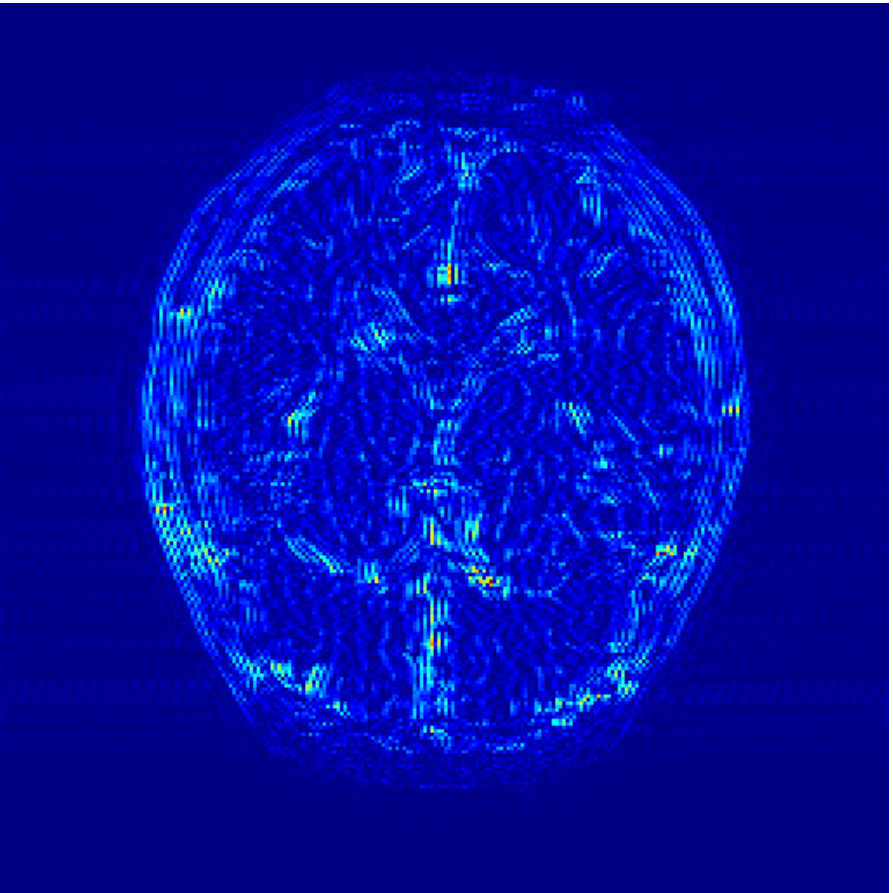}
		\\
		\includegraphics[scale=0.205]{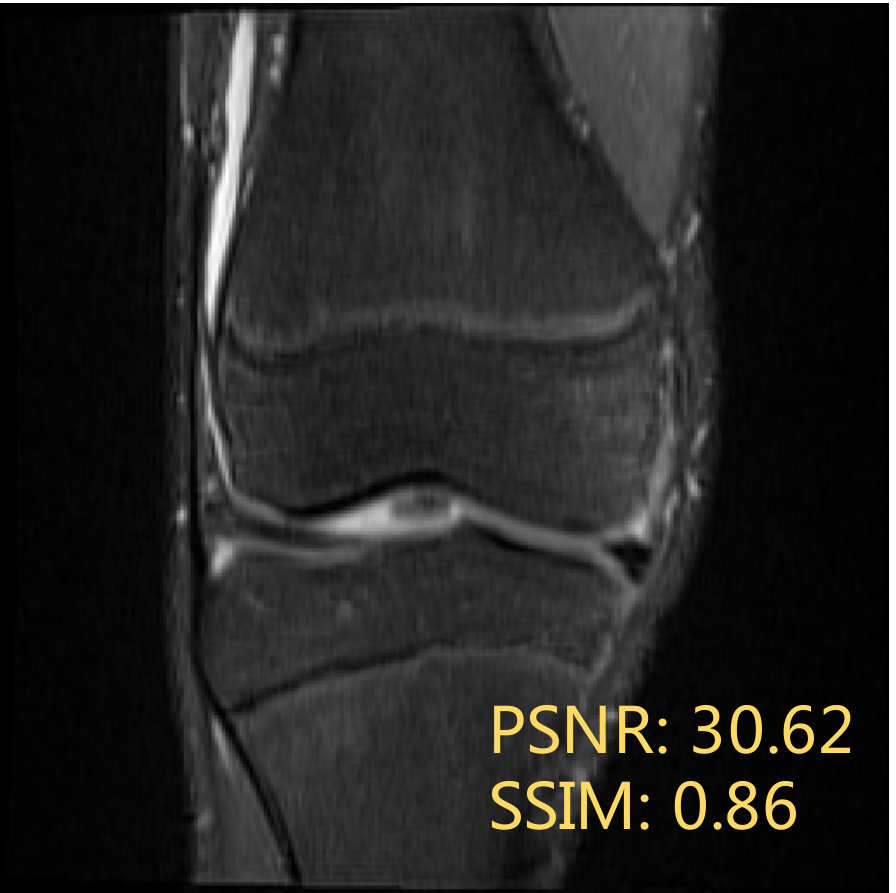}
		\\
		\includegraphics[scale=0.205]{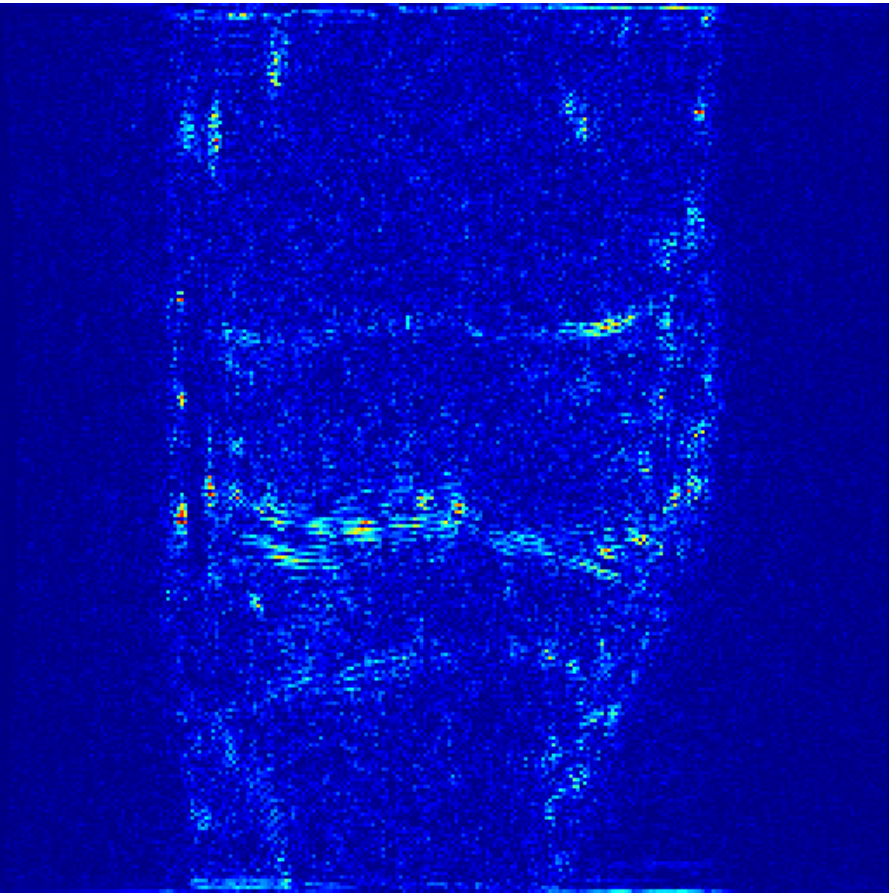}
	\end{minipage}
}
\hfill
    \subfloat[\scriptsize SwinIR \cite{liang2021swinir}]{
	\begin{minipage}[b]{0.1\textwidth}
		\includegraphics[scale=0.205]{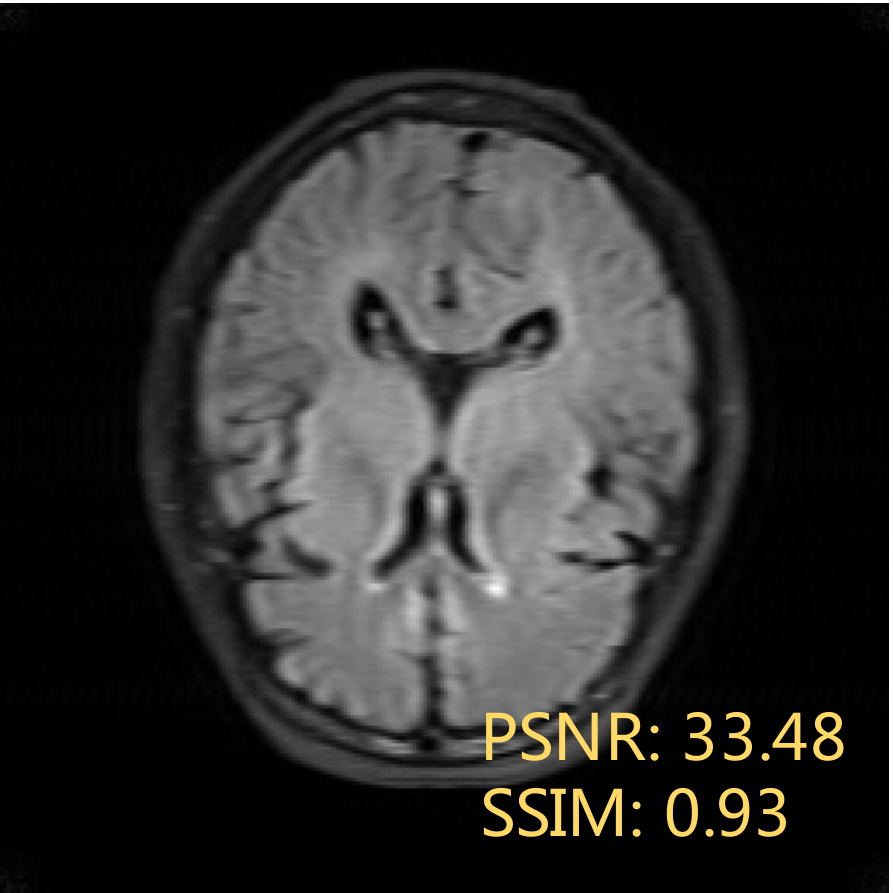} 
		\\
		\includegraphics[scale=0.205]{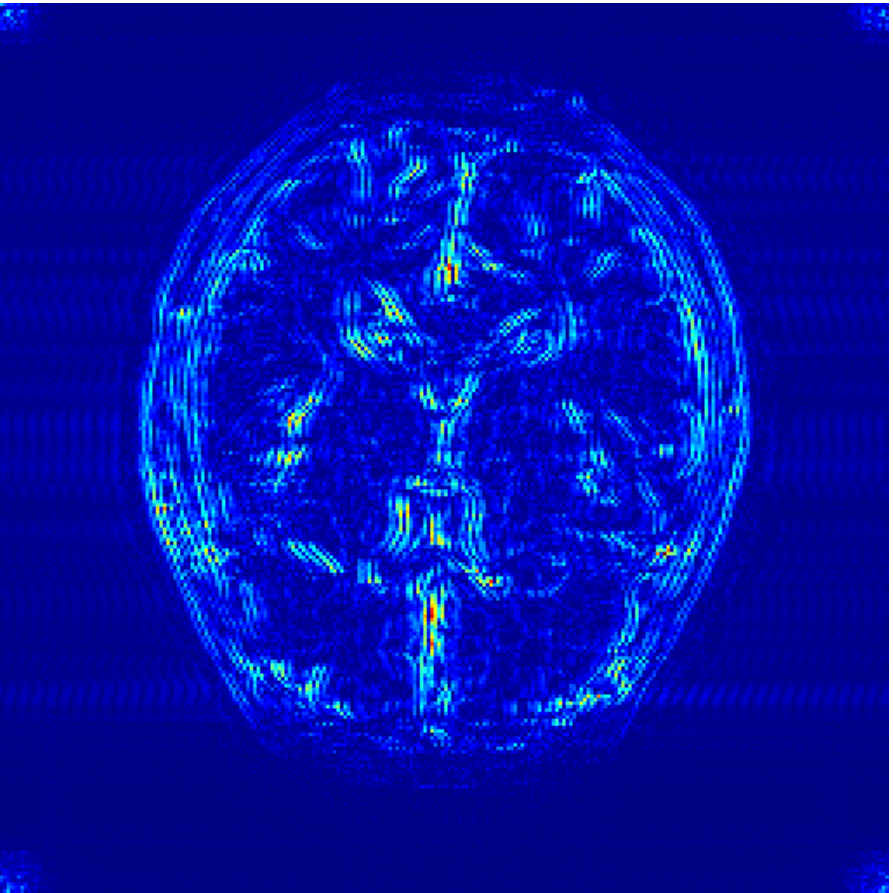}
		\\
		\includegraphics[scale=0.205]{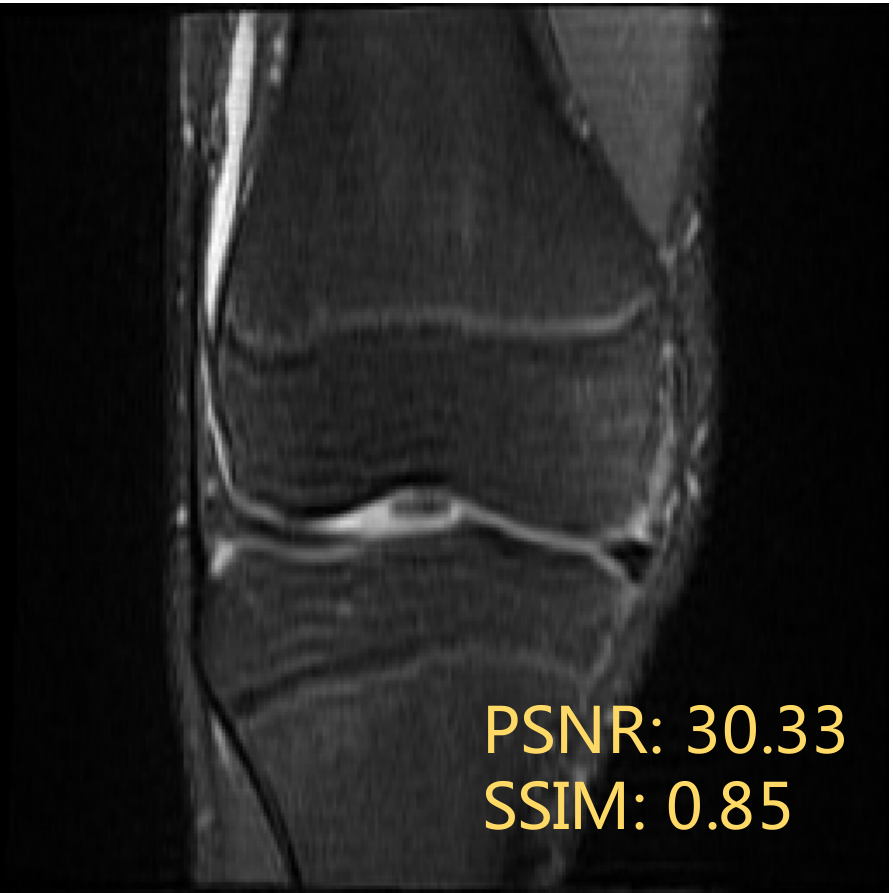}
		\\
		\includegraphics[scale=0.205]{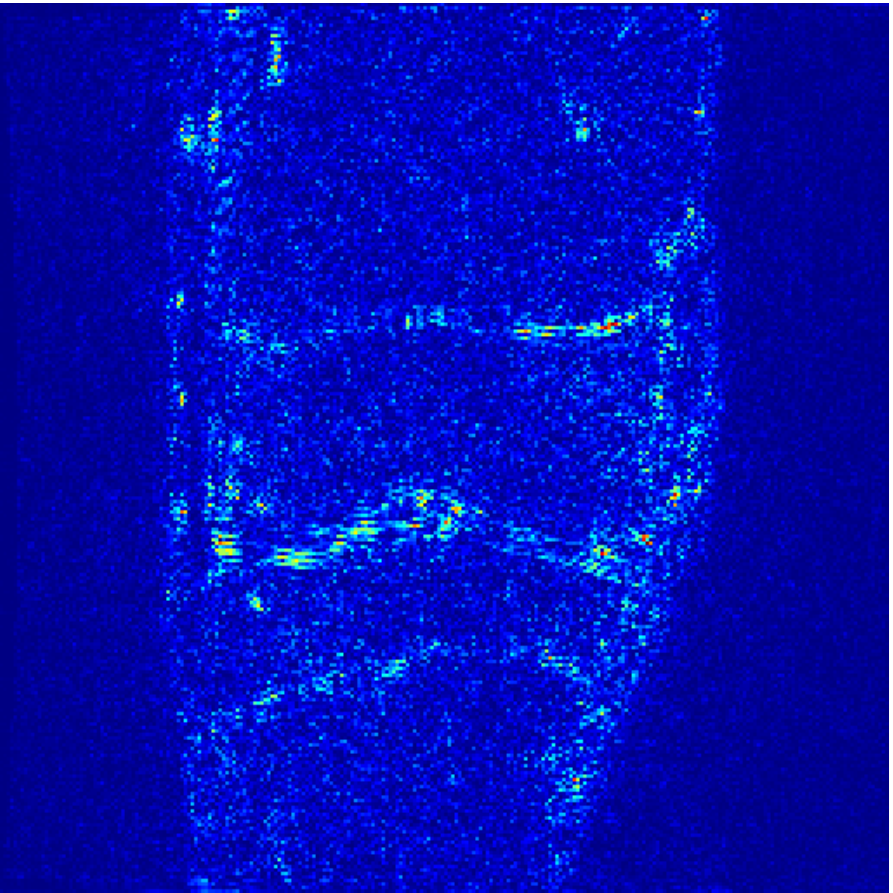}
	\end{minipage}
}
\hfill
    \subfloat[\scriptsize Restormer \cite{zamir2021restormer}]{
	\begin{minipage}[b]{0.1\textwidth}
		\includegraphics[scale=0.205]{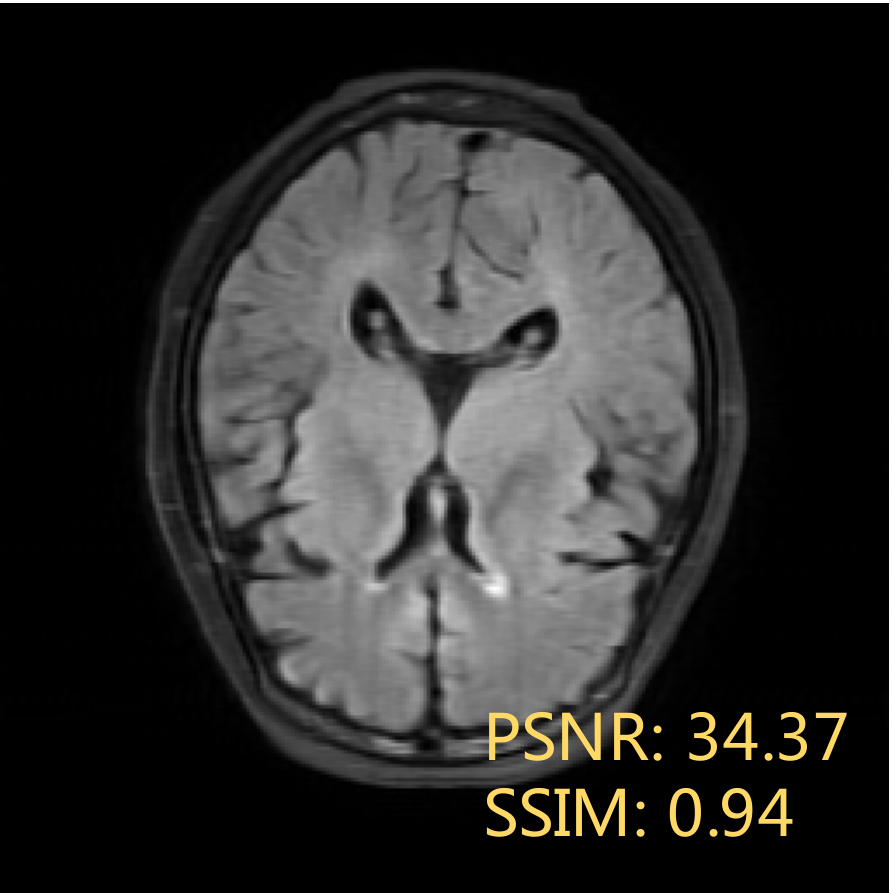} \\
		\includegraphics[scale=0.205]{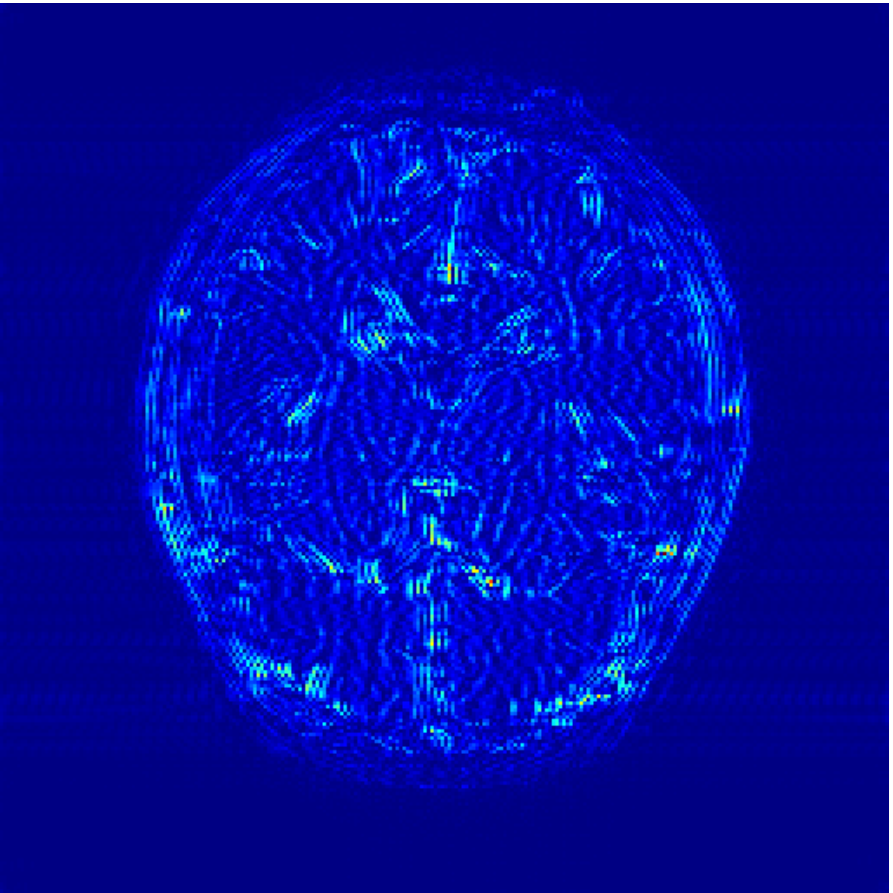}
		\\
		\includegraphics[scale=0.205]{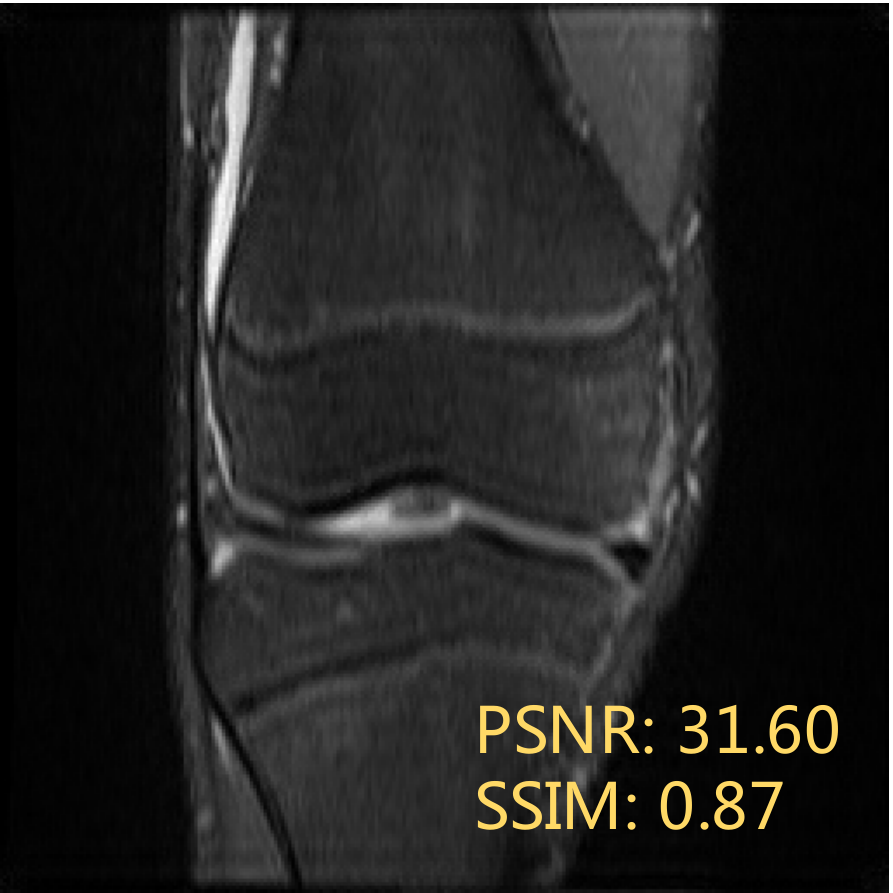}
		\\
		\includegraphics[scale=0.205]{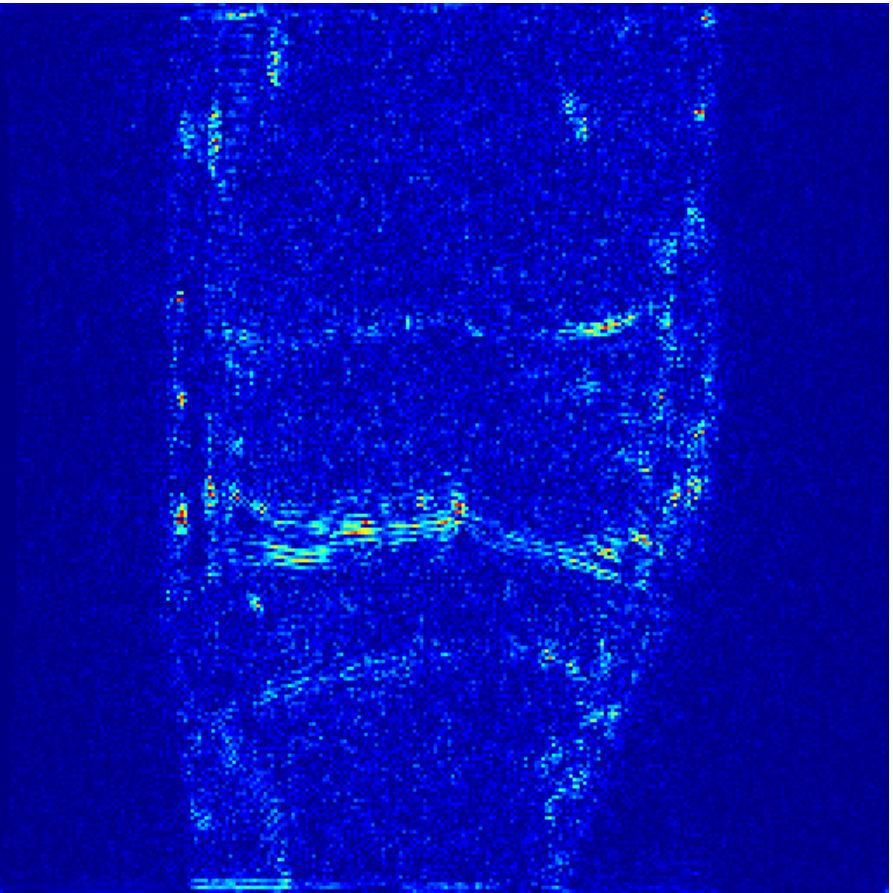}
	\end{minipage}
}
\hfill
    \subfloat[\scriptsize McMRSR]{
	\begin{minipage}[b]{0.1\textwidth}
		\includegraphics[scale=0.205]{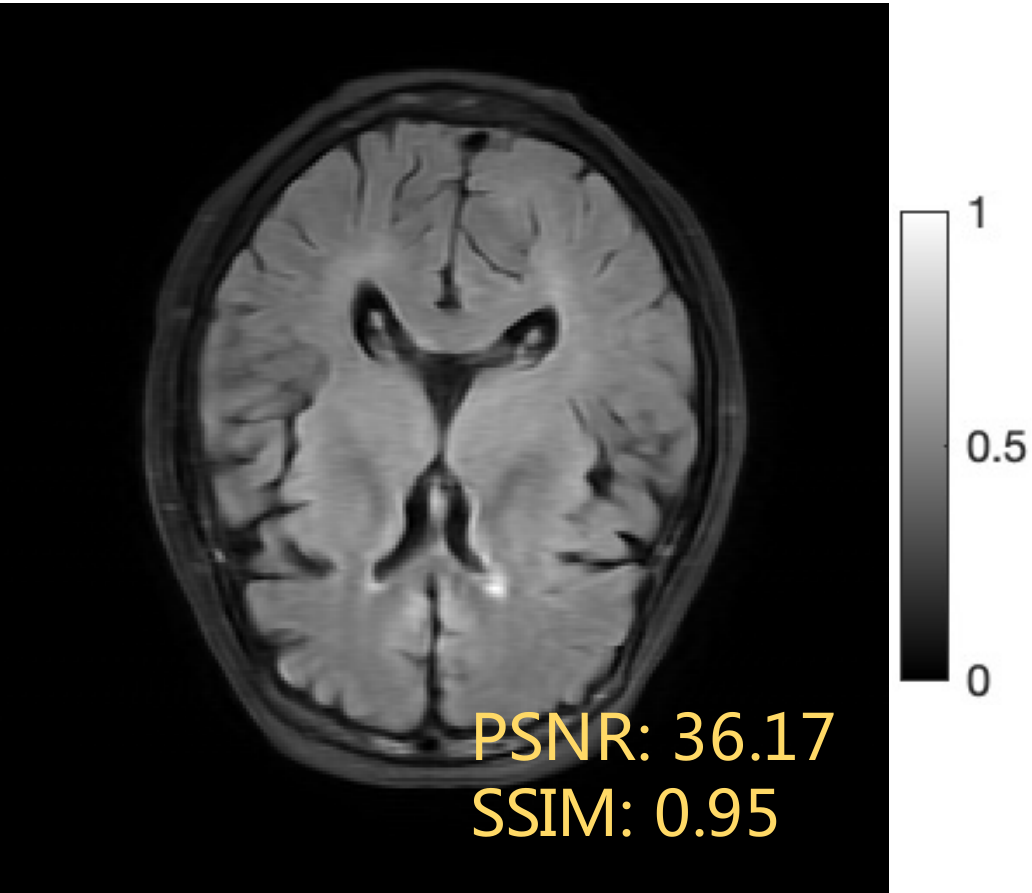} \\
		\includegraphics[scale=0.205]{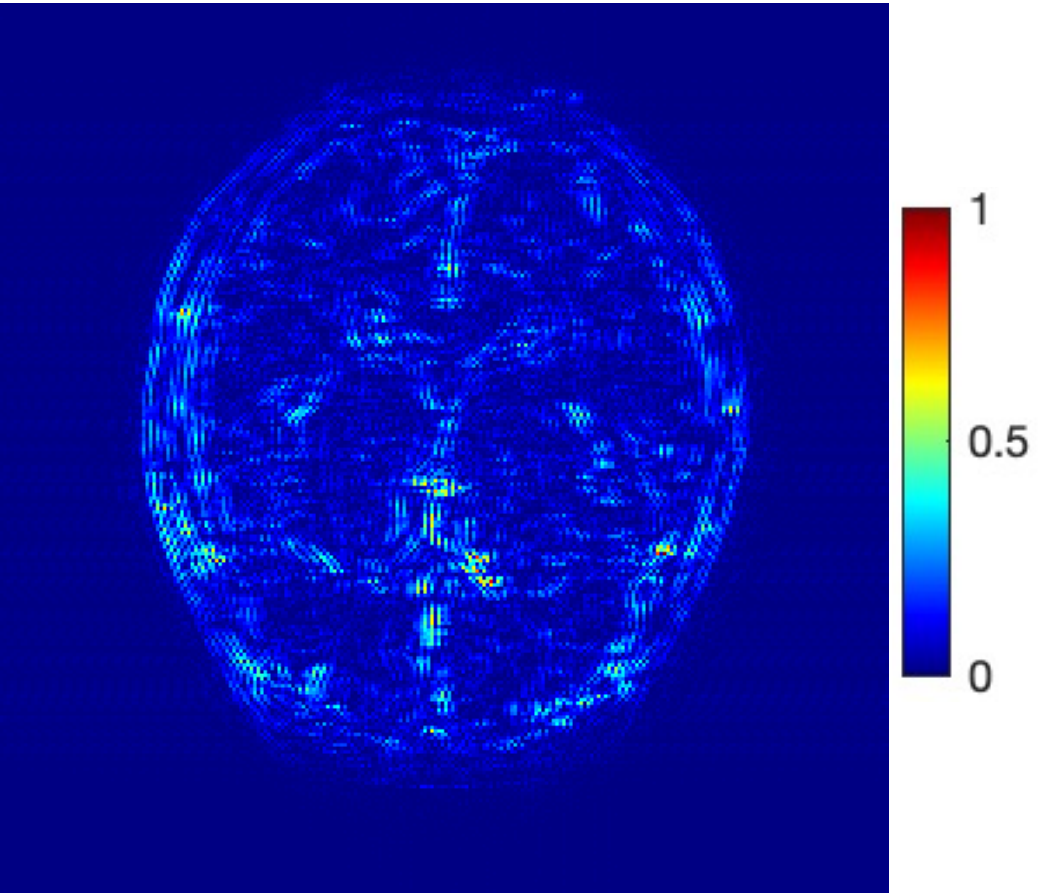}
		\\
		\includegraphics[scale=0.205]{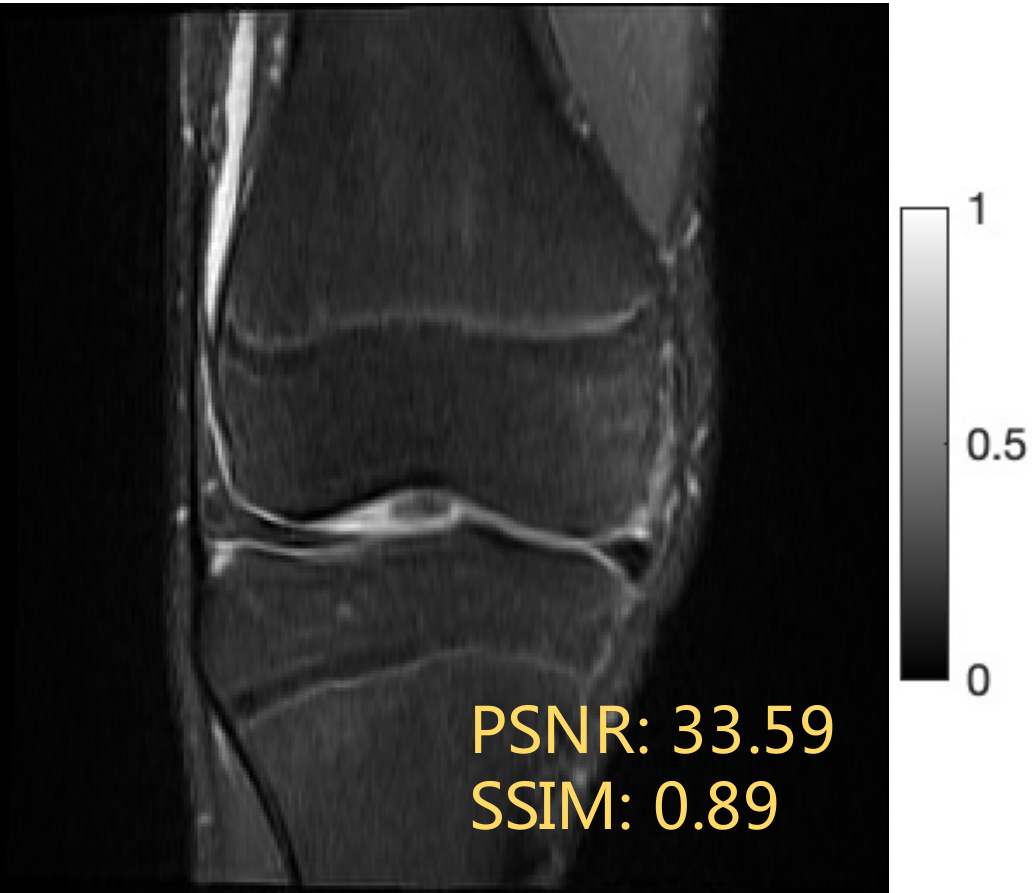}
		\\
		\includegraphics[scale=0.205]{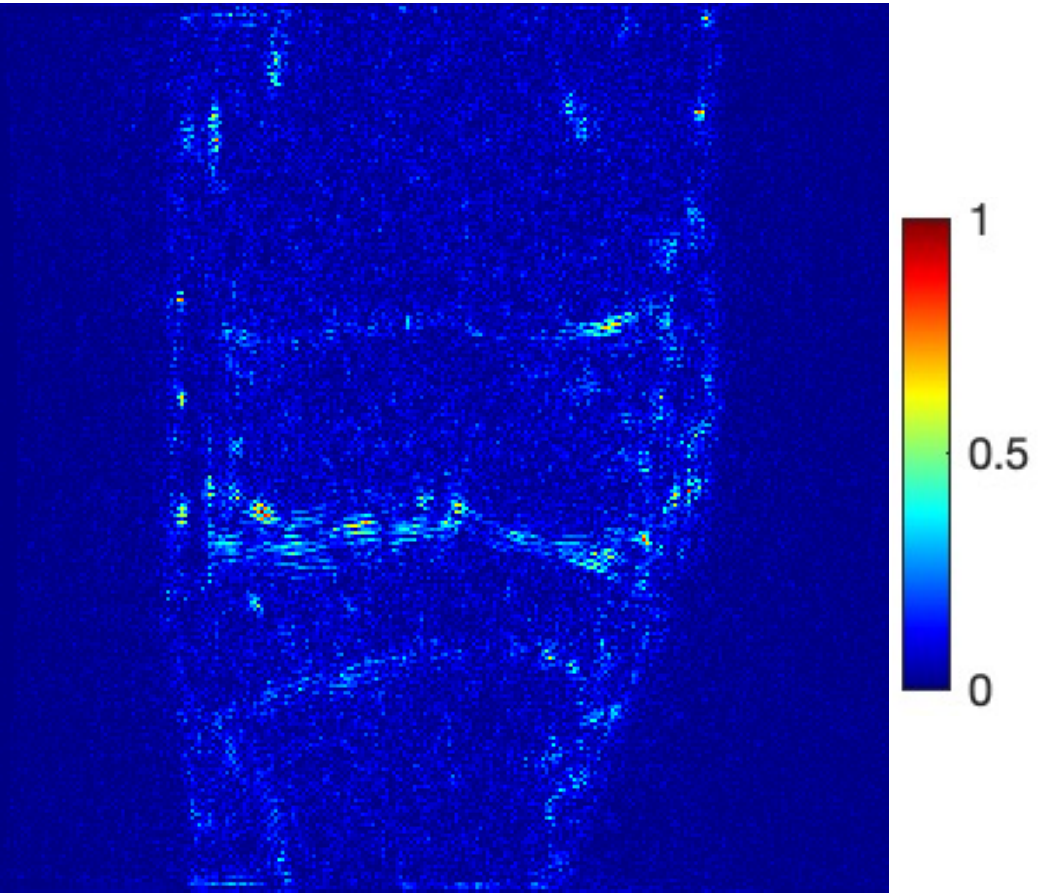}
	\end{minipage}
}
\caption{Qualitative results of different SR reconstruction methods on in-house brain and fastMRI knee datasets with UF=4. The reconstructed images and the corresponding error maps are provided.}
\label{fig:brain_knee}
\end{figure*}

\begin{figure*}[t]
	\centering
	\subfloat[PNSR]{
	\begin{minipage}[b]{0.3\textwidth}
    	\includegraphics[scale=0.33]{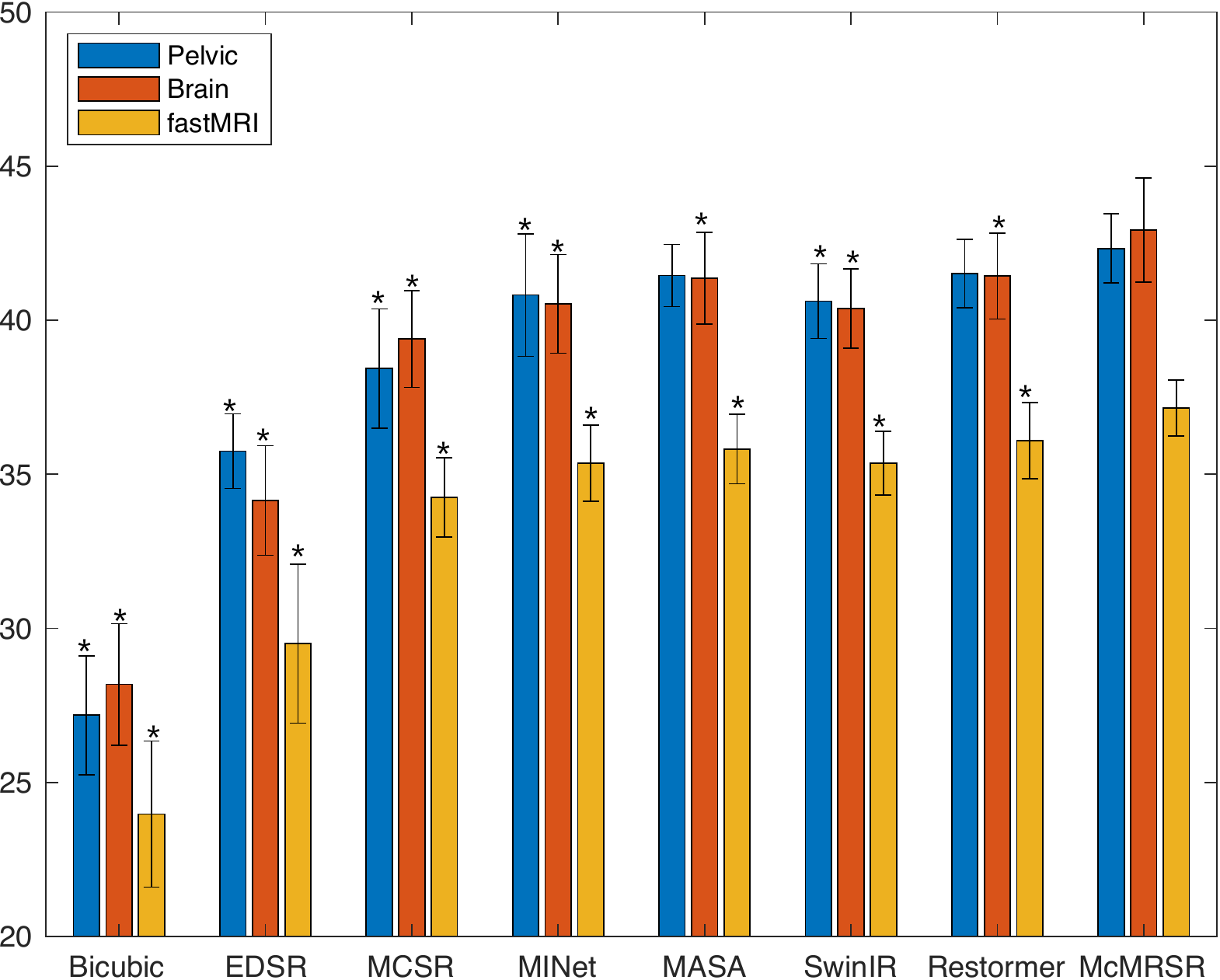}
	\end{minipage}
}
\hfill
	\subfloat[SSIM]{
	\begin{minipage}[b]{0.3\textwidth}
		\includegraphics[scale=0.33]{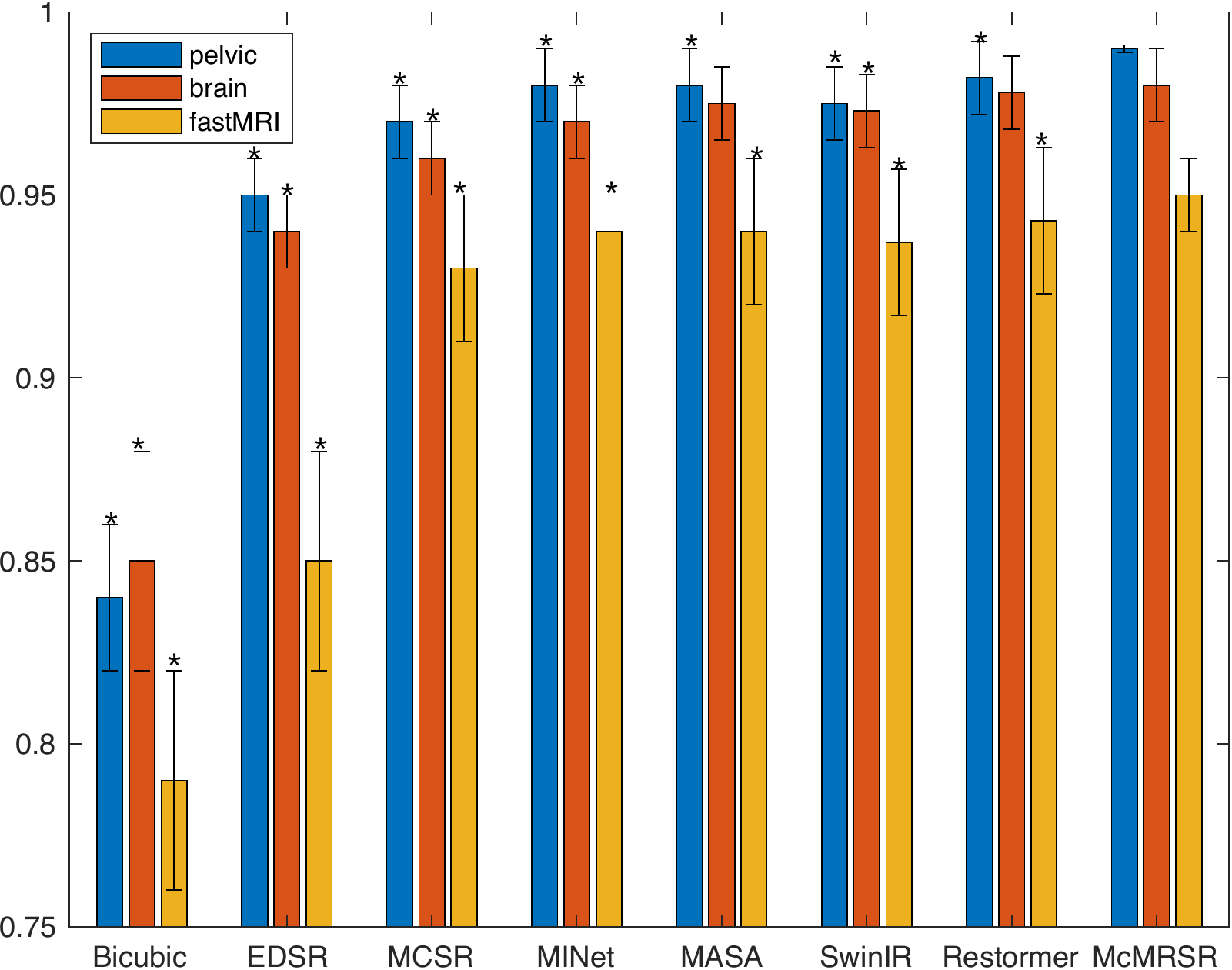} 
	\end{minipage}
}
\hfill
	\subfloat[RMSE($10^{-2}$)]{
	\begin{minipage}[b]{0.3\textwidth}
		\includegraphics[scale=0.33]{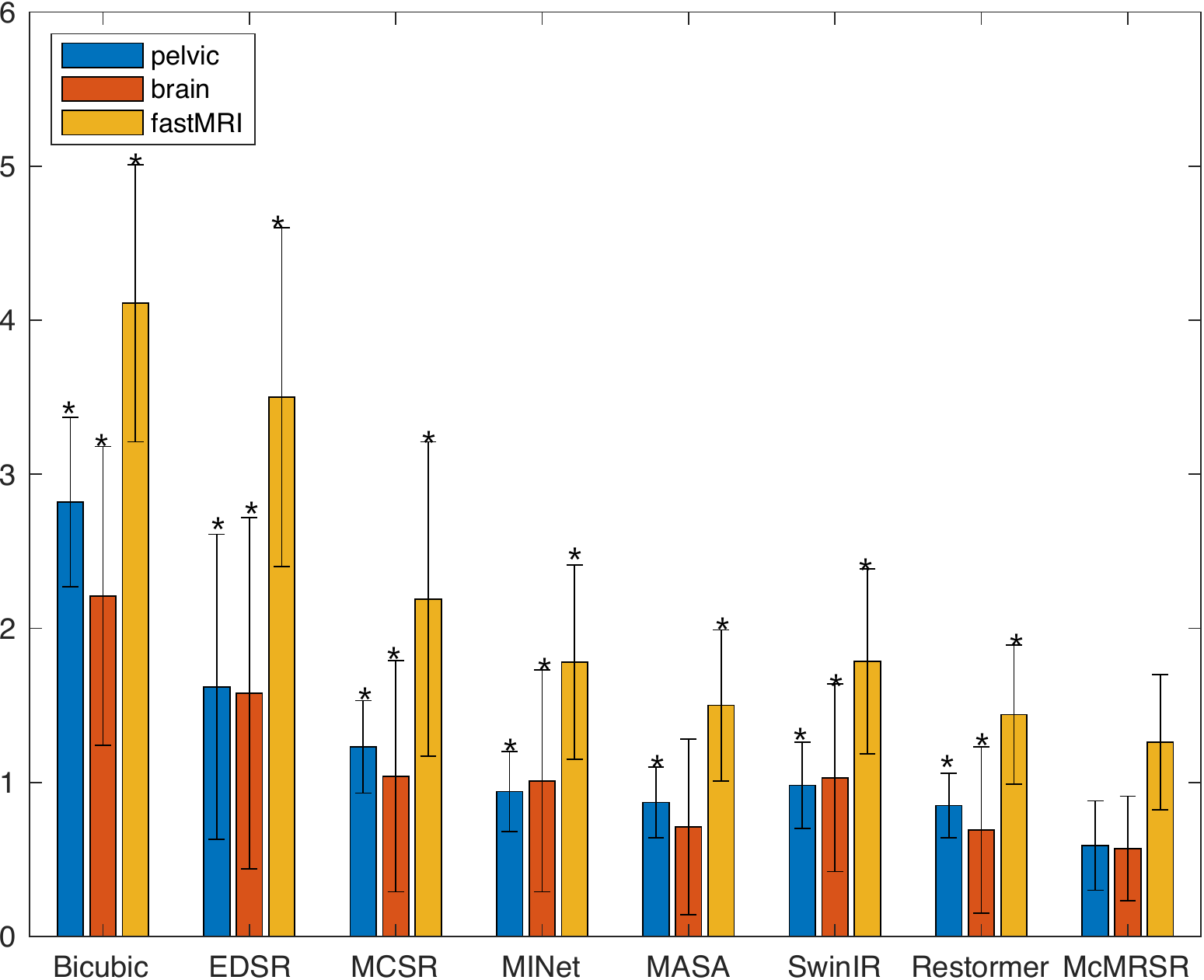} 
	\end{minipage}
}
\vspace{-2mm}
\caption{When UF=2, quantitative metrics results (mean and standard deviation) of different methods with three datasets. $*$ means significant difference between the corresponding method and McMRSR method ($p<$0.01).}
\vspace{-4mm}
\label{fig:metrics_2x}
\end{figure*}

\subsection{Loss Functions}
\subsubsection{Reconstruction Loss}
The L1 pixel loss is utilized as reconstruction loss to improve the overall detail of SR images \cite{ murugesan2019recon }, named as $\mathcal{L}_{rec}$:
\begin{equation}
\mathcal{L}_{r e c}=\mathbb{E}_{\left(\boldsymbol{I}_{S R}, \boldsymbol{I}_{H R}\right)}\left\|\boldsymbol{I}_{S R}-\boldsymbol{I}_{H R}\right\|_{1},
\label{eq:07}
\end{equation}
where $\boldsymbol{I}_{S R}$ denotes reconstructed MR images and $\boldsymbol{I}_{H R}$ denotes original HR MR images.
\subsubsection{\emph{k}-space Data Consistency Loss}
The reconstructed SR images may lose some frequency domain information in the original HR images.
We introduce the $k$-space data consistency \cite{ zhou2020dudornet } to militate this. 
%
Specifically,
$\boldsymbol{K}_{SR}$ and $\boldsymbol{K}_{HR}$ denotes the fast Fourier transform of $\boldsymbol{I}_{SR}$ and $\boldsymbol{I}_{HR}$. 
Then, the sampling judgment is performed using $R_{l r}$. If the coefficients in $\boldsymbol{K}_{SR}$ have been sampled, they are replaced with those in $\boldsymbol{K}_{HR}$, otherwise they remain unchanged. The final fidelity of the $k$-space image is obtained, 
this process can be expressed as:
\begin{eqnarray}
\boldsymbol{K}_{D C}[a, b] = \begin{cases} \boldsymbol{K}_{S R}[a, b] & \mbox{if }(a, b) \notin R_{l r}\\
\frac{\boldsymbol{K}_{S R}[a, b]+n \boldsymbol{K}_{H R}[a, b]}{1+n} & \mbox{if }(a, b) \in R_{l r} \end{cases},
\label{eq:08}
\end{eqnarray}
where $R_{l r}$ is defined as the LR mask, $n\geq0$ is the noise level (here $n$ is set to infinity), and $[a,b]$ is the matrix indexing operation. We use mean squared error (MSE) to measure the error between $\boldsymbol{K}_{DC}$ and $\boldsymbol{K}_{HR}$ as:
\begin{equation}
\mathcal{L}_{d c}=\mathbb{E}_{\left(\boldsymbol{K}_{D C}, \boldsymbol{K}_{H R}\right)}\left\|\boldsymbol{K}_{D C}-\boldsymbol{K}_{H R}\right\|_{2}.
\label{eq:09}
\end{equation}
To the end, the full objective of the McMRSR network is defined as:
\begin{equation}
\mathcal{L}_{f u l l}=\lambda_{r e c} \mathcal{L}_{r e c}+\lambda_{d c} \mathcal{L}_{d c}.
\label{eq:10}
\end{equation}
We set $\lambda_{rec}=1$ and $\lambda_{dc}=0.0001$ so that the magnitude of different loss terms can be balanced into similar scales, making their contributions reasonable.

\section{Experiments}
\subsection{Datasets and Baselines}
Three datasets are utilized to evaluate the proposed McMRSR network, including two in-house datasets of pelvic and brain and one public fastMRI \cite{ zbontar2018fastmri } dataset, as shown in \cref{tab_data}.
All the complex-valued images are reshaped into the matrix size of 256$\times$256 by cropping the $k$-space. 
We adopt a commonly used downsampling treatment, which is implemented in the frequency domain \cite{lyu2020multi,lyu2020mri}. Specifically, we first converted the original image of size 256$\times$256 into the $k$-space. Then, only data in the central low-frequency region are kept, and all the peripheral data points are zeroed out. For the down-sampling factors 2$\times$ and 4$\times$, the central 25$\%$ and 6.25$\%$ data points are kept. Finally, we used the inverse Fourier transform to convert the modified data into the image domain to produce the LR image.

\begin{table}[h]
\scriptsize
\caption{Three datasets used to evaluate the proposed McMRSR.}
\vspace{-1mm}
\begin{tabular}{p{50pt}|p{48pt}|p{48pt}|p{48pt}}
\hline
\hline
 \makecell[l]{Datasets } & \makecell[c]{Pelvic}  & \makecell[c]{Brain} & \makecell[c]{fastMRI \cite{ zbontar2018fastmri }}  \\  
 \hline
  \makecell[l]{Reference}  & \makecell[c]{T1} & \makecell[c]{T1} & \makecell[c]{PD} \\
 \hline
 \makecell[l]{Target}  & \makecell[c]{T2} & \makecell[c]{T2-FLAIR} & \makecell[c]{FS-PD} \\
 \hline
  \makecell[l]{Train$\backslash$Valid$\backslash$Test}  & \makecell[c]{1280$\backslash$320$\backslash$320} & \makecell[c]{513$\backslash$125$\backslash$125} & \makecell[c]{320$\backslash$80$\backslash$80} \\
\hline
\hline
\end{tabular}
\vspace{-3mm}
\label{tab_data}
\end{table}

\begin{table*}[h]
\scriptsize
\centering
\caption{Quantitative metrics results (mean and standard deviation) on different datasets with 4$\times$ enlargement scale, in terms of PSNR, SSIM, and RMSE ($\times10^{-2}$). \textbf{Bold} is the best results. All comparison methods have significant difference with our method ($p<$0.01). }
\vspace{-2mm}
\begin{tabular}{p{75pt}|p{33pt}|p{32pt}|p{32pt}|p{33pt}|p{32pt}|p{32pt}|p{33pt}|p{32pt}|p{32pt}}   
\hline
\hline
\makecell[l]{Dataset} & \multicolumn{3}{c|}{Pelvic} &\multicolumn{3}{c|}{Brain} &\multicolumn{3}{c}{fastMRI \cite{ zbontar2018fastmri }}\\  
\hline
\makecell[l]{Metrics} 
& \makecell[c]{PSNR} & \makecell[c]{SSIM} &\makecell[c]{RMSE}  
& \makecell[c]{PSNR} & \makecell[c]{SSIM} &\makecell[c]{RMSE}  
& \makecell[c]{PSNR} & \makecell[c]{SSIM} &\makecell[c]{RMSE} \\
\hline
\hline
\makecell[l]{Bicubic} 
& \makecell[c]{24.38(2.30)} & \makecell[c]{0.76(0.03)} &\makecell[c]{6.22(2.11)}  
& \makecell[c]{22.30(2.78)} & \makecell[c]{0.74(0.02)} &\makecell[c]{8.08(2.53)}  
& \makecell[c]{19.15(2.37)} & \makecell[c]{0.64(0.03)} &\makecell[c]{7.23(2.16)} \\
\hline
\makecell[l]{EDSR (CVPR2017) \cite{ lim2017enhanced }} 
& \makecell[c]{27.39(1.55)} & \makecell[c]{0.85(0.02)} &\makecell[c]{4.32(1.75)}  
& \makecell[c]{26.01(2.30)} & \makecell[c]{0.84(0.02)} &\makecell[c]{6.79(2.32)}  
& \makecell[c]{24.26(1.62)} & \makecell[c]{0.69(0.02)} &\makecell[c]{6.42(2.13)} \\
\hline
\makecell[l]{MCSR (TMI2020) \cite{ lyu2020multi }} 
& \makecell[c]{32.12(1.01)} & \makecell[c]{0.92(0.01)} &\makecell[c]{2.61(1.00)}  
& \makecell[c]{32.09(1.95)} & \makecell[c]{0.93(0.01)} &\makecell[c]{4.93(1.46)}  
& \makecell[c]{28.09(1.25)} & \makecell[c]{0.82(0.03)} &\makecell[c]{3.25(1.03)} \\
\hline
\makecell[l]{MINet (MICCAI2021) \cite{ feng2021multi }} 
& \makecell[c]{34.41(0.85)} & \makecell[c]{0.94(0.01)} &\makecell[c]{1.82(1.27)}  
& \makecell[c]{34.32(1.08)} & \makecell[c]{0.94(0.01)} &\makecell[c]{3.05(1.75)}  
& \makecell[c]{30.58(1.38)} & \makecell[c]{0.86(0.02)} &\makecell[c]{2.91(0.99)} \\
\hline
\makecell[l]{MASA (CVPR2021) \cite{ lu2021masa }} 
& \makecell[c]{34.86(1.14)} & \makecell[c]{0.94(0.02)} &\makecell[c]{1.59(1.46)}  
& \makecell[c]{34.79(1.06)} & \makecell[c]{0.94(0.01)} &\makecell[c]{2.57(1.62)}  
& \makecell[c]{30.97(1.14)} & \makecell[c]{0.86(0.03)} &\makecell[c]{2.70(0.90)} \\
\hline
\makecell[l]{SwinIR (ICCV2021) \cite{liang2021swinir}} 
& \makecell[c]{33.92(1.09)} & \makecell[c]{0.93(0.01)} &\makecell[c]{2.10(1.52)}  
& \makecell[c]{34.08(1.78)} & \makecell[c]{0.93(0.02)} &\makecell[c]{3.37(1.66)}  
& \makecell[c]{30.36(1.34)} & \makecell[c]{0.85(0.02)} &\makecell[c]{2.98(1.09)} \\
\hline
\makecell[l]{Restormer (arXiv) \cite{zamir2021restormer}} 
& \makecell[c]{34.91(1.18)} & \makecell[c]{0.94(0.02)} &\makecell[c]{1.48(1.49)}  
& \makecell[c]{34.73(1.89)} & \makecell[c]{0.94(0.03)} &\makecell[c]{2.54(2.07)}  
& \makecell[c]{31.09(1.05)} & \makecell[c]{0.86(0.02)} &\makecell[c]{2.59(1.48)} \\
\hline
\makecell[l]{McMRSR (Ours)} 
& \makecell[c]{\textbf{36.23(1.07)}} & \makecell[c]{\textbf{0.96(0.01)}} &\makecell[c]{\textbf{1.09(0.89)}}  
& \makecell[c]{\textbf{36.07(0.92)}} & \makecell[c]{\textbf{0.95(0.01)}} &\makecell[c]{\textbf{1.73(1.08)}}  
& \makecell[c]{\textbf{33.28(0.97)}} & \makecell[c]{\textbf{0.90(0.02)}} &\makecell[c]{\textbf{1.82(0.85)}} \\
\hline
\hline
\end{tabular}
\label{tab_metrics}
\end{table*}


\begin{table*}[h!]
\scriptsize
\centering
\caption{Ablation study on different variant model under fastMRI dataset with 4$\times$ enlargement scale. The best quantitative metrics results is marked in \textbf{bold}. There has a significant difference between variant models and McMRSR model ($p<$0.01). RMSE ($\times10^{-2}$).}
\vspace{-2mm}
\begin{tabular}{p{48pt}|p{48pt}|p{37pt}|p{38pt}|p{38pt}|p{37pt}|p{38pt}|p{38pt}}   
\hline
\hline
\makecell[l]{\multirow{2}*{Variant}} & \multicolumn{4}{c|}{Modules} &\multicolumn{3}{c}{Metrics} \\  
\cline{2-8}
& \makecell[c]{Reference-Based} & \makecell[c]{Multi-Scale} &\makecell[c]{CM}  
& \makecell[c]{MAB} & \makecell[c]{PSNR} &\makecell[c]{SSIM} &\makecell[c]{RMSE}  \\
\hline
\hline
\makecell[l]{$w/o$ reference} 
& \makecell[c]{$\times$} & \makecell[c]{\checkmark} &\makecell[c]{\checkmark}  
& \makecell[c]{\checkmark} 
& \makecell[c]{29.05(1.24)} &\makecell[c]{0.83(0.02)} 
&\makecell[c]{3.23(0.97)} \\
\hline
\makecell[l]{$w/o$ multi-scale} 
& \makecell[c]{\checkmark} & \makecell[c]{$\times$} &\makecell[c]{\checkmark}  
& \makecell[c]{\checkmark} 
& \makecell[c]{30.24(1.12)} &\makecell[c]{0.85(0.03)} 
&\makecell[c]{3.04(0.93)} \\
\hline
\makecell[l]{$w/o$ CM} 
& \makecell[c]{\checkmark} & \makecell[c]{\checkmark} &\makecell[c]{$\times$}  
& \makecell[c]{\checkmark} 
& \makecell[c]{31.13(1.18)} &\makecell[c]{0.86(0.02)} 
&\makecell[c]{2.66(0.79)} \\
\hline
\makecell[l]{$w/o$ MAB} 
& \makecell[c]{\checkmark} & \makecell[c]{\checkmark} &\makecell[c]{\checkmark}  
& \makecell[c]{$\times$} 
& \makecell[c]{30.56(1.03)} &\makecell[c]{0.85(0.03)} 
&\makecell[c]{2.95(0.83)} \\
\hline
\makecell[l]{McMRSR} 
& \makecell[c]{\checkmark} & \makecell[c]{\checkmark} &\makecell[c]{\checkmark}  
& \makecell[c]{\checkmark} &
\makecell[c]{\textbf{33.28(0.97)}}   & \makecell[c]{\textbf{0.90(0.02)}}  & \makecell[c]{\textbf{1.82(0.85)}}\\
\hline
\hline
\end{tabular}
\label{tab_ab}
\vspace{-2mm}
\end{table*}

We compared our McMRSR with several recent state-of-the-art methods, including a single-contrast SR method: EDSR \cite{ lim2017enhanced }, three multi-contrast SR methods: MCSR \cite{ lyu2020multi }, MINet\cite{ feng2021multi }, MASA \cite{ lu2021masa }, and two transformer-based SR methods: SwinIR \cite{liang2021swinir}, Restormer \cite{zamir2021restormer}. Note that we concatenate the reference contrast and the target contrast as input for SwinIR and Restormer.

\subsection{Implementation Details}
Our proposed McMRSR is implemented in PyTorch with NVIDIA Tesla V100 GPUs (4$\times$16GB). The Adam \cite{ kingma2014adam } optimizer is adopted for model training with the learning rate of $10^{-4}$ and epochs of 200. The performance of the SR reconstruction is evaluated by peak-to-noise-ratio (PSNR), structural similarity index (SSIM), and root mean squared error (RMSE) metrics. 
In addition, we use ranksum to calculate whether there is a significant difference between McMRSR and other comparison methods ($p<$0.01).
The upsampling factors are set to 2$\times$ and 4$\times$, respectively. 

\subsection{Qualitative Results}
\cref{fig:pelvic} provides the reconstruction results and the corresponding error maps of pelvic images when the UF=4. The predominant texture in the error map means worse reconstruction quality. As can be observed, the reconstructed SR images from the multi-contrast methods are significantly better than those from the single-contrast EDSR approach,
demonstrating the effectiveness of complementary information embedded in multi-contrast images in the task of MRSR.  
%
More importantly, the SR image reconstructed by our McMRSR can better recover the uterine part and eliminate blurring edges, thanks to the proposed contextual matching and aggregation schemes.   

In order to demonstrate the generalization capability and robustness of our method, we further conducted experiments on brain and fastMRI datasets, and the visual results are shown in \cref{fig:brain_knee}. Similarly, we can see that our method is able to restore more anatomical details in both brain and knee datasets.
Moreover, it can effectively handle various noises or artifacts in the images, thanks to the long-range contextual information captured by the transformers equipped in our model.

\subsection{Quantitative Results}

\cref{fig:metrics_2x} reports the metrics scores with different datasets under 2$\times$ enlargement. As can be seen, our model yields the best results in terms of all metrics.
We further calculate the results in terms of all metrics for each method under 4$\times$ enlargement, as shown in \cref{tab_metrics}. Although it is more challenging to restore SR images under 4$\times$ enlargement than 2$\times$, our method consistently outperforms existing methods with the best metrics scores. 

\subsection{Ablation Study}
In this section, we demonstrate the effectiveness of the key components of McMRSR through ablation studies. The ablation studies are performed using fastMRI dataset with UF=4. 
In order to verify if Swin Transformer can effectively extract the deep features of images and better recover SR images,
we design a single-contrast variant model using Swin Transformer, named as $w/o$ reference. 
This variant model does not perform multi-scale context matching and aggregation of reference features, but only perform upsampling operation on features in target LR. 
To verify that context matching and aggregation at multi-scale is superior to single-scale, we design a single-scale variant model, named as $w/o$ multi-scale. 
To verify the contribution of context matching and MAB in the model, we further designed variant networks without context matching (CM) for reference features, named as $w/o$ CM and without MAB for upsampling, named as $w/o$ MAB. 
The quantitative metrics results of these variant models are shown in \cref{tab_ab}.

As can be seen, the results of $w/o$ reference are still better than the EDSR (in \cref{tab_metrics}), indicating that transformers are able to extract more representative features with rich long-range dependencies for better reconstruction. 
%
More importantly, quantitative metrics scores of McMRSR are better than those of other multi-contrast variant models. 
This indicates the proposed multi-scale context matching and aggregation schemes are effective and capable of providing more reference features than previous approaches. 
Our context matching scheme ensures that the reference features at each scale contain the features most relevant to the target LR. 
In addition, MAB drives the target LR to maximize the use of multi-scale matched reference features during the upsampling process.

\subsection{Limitation and Future Work}
Here, we discuss the limitations and potential future works of this study.
First, the LR-HR multi-contrast image pairs need to be co-registered in advance, which is tedious and time-consuming. 
In the future, we shall work on design a multi-task framework to simultaneously preform registration and SR reconstruction.
Second, although our model achieves state-of-the-art performance, the reconstructed MR images still contain some artifacts, which may lead to incorrect diagnoses. 
In this regard, we shall further explore the fundamental limits of learning techniques for SR reconstruction and strive to design better approaches to tackle these artifacts.


\section{Conclusion }
We present a novel transformer-empowered multi-scale contextual matching and aggregation network for multi-contrast MRI SR reconstruction. 
Our model can make full use of information embedded in the reference image and reconstruct the target image with close quality to the original target HR on three representative datasets with both 2$\times$ and 4$\times$ upsampling scales.
Specifically, our method provides sufficient complementary information for target LR features by harnessing contextual matching and aggregating the reference features at different scales. 
Experimental results show that our method is superior to the existing multi-contrast MRI SR methods and has potential to be used in clinical practice.

\textbf{Acknowledgments.} This work was supported by the National Natural Science Foundation of China under Grant 61902338 and Hong Kong Research Grants Council under General Research Fund 15205919.

{\small
\bibliographystyle{ieee_fullname}
\bibliography{egbib}
}

\end{document}